\documentclass[12pt,preprint]{aastex}

\usepackage{amsmath}

\usepackage{epsf}
\usepackage{cancel}

\usepackage{color}
\shortauthors{Almgren,Bell,Lijewski,Lukic,Van Andel}
\shorttitle{Nyx Code Paper}

\newcommand{\ubold}{U}
\newcommand{\Ub}{{\bf U}}
\newcommand{\Fb}{{\bf F}}

\newcommand{\eb}{{\bf e}}
\newcommand{\Dt}{\Delta t}
\def\ellp {{\ell^\prime}}

\newcommand{\sfrac}[2]{\mathchoice
  {\kern0em\raise.5ex\hbox{\the\scriptfont0 #1}\kern-.15em/
   \kern-.15em\lower.25ex\hbox{\the\scriptfont0 #2}}
  {\kern0em\raise.5ex\hbox{\the\scriptfont0 #1}\kern-.15em/
   \kern-.15em\lower.25ex\hbox{\the\scriptfont0 #2}}
  {\kern0em\raise.5ex\hbox{\the\scriptscriptfont0 #1}\kern-.2em/
   \kern-.15em\lower.25ex\hbox{\the\scriptscriptfont0 #2}}
  {#1\!/#2}}

\def\half   {\frac{1}{2}}
\def\myhalf {\sfrac{1}{2}}

\def\gb {\mathbf{g}}
\def\xb {\mathbf{x}}
\def\ub {\mathbf{u}}
\def\nph {{n+\myhalf}}
\def\dt {\Delta t}
\def\dx {\Delta x}

\newcommand{\Cplusplus}{{\rmfamily C\raise.22ex\hbox{\small ++} }}
\newcommand{\Code}[1]{\texttt{#1}}

\begin{document}

\title{Nyx: A Massively Parallel AMR Code for Computational Cosmology}

\shorttitle{Nyx}
\shortauthors{Almgren et al.}

\author{Ann~S.~Almgren\altaffilmark{1},
        John~B.~Bell\altaffilmark{1},
        Mike~J.~Lijewski\altaffilmark{1},
        Zarija~Luki\'c\altaffilmark{2},
        Ethan~Van~Andel\altaffilmark{3}}

\altaffiltext{1}{Center for Computational Sciences and Engineering,
                 Lawrence Berkeley National Laboratory,
                 Berkeley, CA 94720}

\altaffiltext{2}{Computational Cosmology Center,
                 Lawrence Berkeley National Laboratory,
                 Berkeley, CA 94720}

\altaffiltext{3}{Mathematics Department,
                 UC Berkeley,
                 Berkeley, CA 94720}


\clearpage
\begin{abstract}

We present a new N-body and gas dynamics code, called Nyx, for large-scale cosmological simulations. 
Nyx follows the temporal evolution of a system of discrete dark matter particles gravitationally coupled 
to an inviscid ideal fluid in an expanding universe.  The gas is advanced in
an Eulerian framework with block-structured adaptive mesh refinement (AMR); a particle-mesh (PM) scheme 
using the same grid hierarchy is used to solve for self-gravity and advance
the particles.  Computational results demonstrating 
the validation of Nyx on standard cosmological test problems,
and the scaling behavior of Nyx to 50,000 cores, are presented. 
\end{abstract}

\section{Introduction}\label{sec:intro}

Understanding how the distribution of matter in the Universe evolves throughout cosmic history is one of 
the major goals of cosmology. Properties of the large-scale structure depend strongly on 
a cosmological model, as well as the initial conditions in that model. 
Thus, in principle, we can determine the cosmological parameters of our Universe by matching the observed 
distribution of matter to the best-fit calculated model universe. This is much easier 
said then done. Observations, with the exception of gravitational lensing, cannot directly access 
the dominant form of matter, the dark matter. Instead they reveal information about baryons, 
which in some biased way trace the dark matter. On the theory side, equations describing the 
evolution of density perturbations (e.g.~\citet{peebles1980}) can be easily solved 
only when the amplitude of perturbations 
is small. However, in most systems of interest, the local matter density, 
$\rho,$ can be several orders of magnitude greater than the average density, 
$\rho_0,$ making the density contrast, $\delta \equiv (\rho - \rho_0)/\rho_0,$ 
much greater then unity. 

Due to this non-linearity, and the uncertainty in the initial conditions of the Universe, 
a single ``heroic" simulation is not sufficient to resolve most of the cosmological questions
of interest; the ability to run many simulations of different cosmologies, with different 
approximations to the governing physics, is imperative.  Semi-analytic models 
(for example \citet{white1991}, in the context of galaxy formation) or scaling relations 
(e.g.~\citet{smith2003}) present an interesting alternative to 
expensive simulations, but they must also be validated or calibrated by simulations in 
order to be trustworthy. Finally, besides being an invaluable theorist's tool, 
simulations play a crucial role in bridging the gap between theory and observation, 
something that is becoming more and more important as large sky surveys are starting to collect data. 

To match the capabilities of future galaxy/cluster surveys, cosmological simulations must be able to 
simulate volumes $\sim$1 Gpc on a side while resolving scales on the order of $\sim$10 kpc or even smaller.  
Clearly, in the context of Eulerian gas dynamics, this requires the ability to run simulations with 
multiple levels of refinement that follow the formation of relevant objects. 
To resolve $L_*$ galaxies with $\sim$100 mass elements (particles) per galaxy in 
simulations of this scale, one needs tens of billions of particles. 
Such systems, with crossing times of a million years or less, must be
evolved for $\sim$10 billion years.
On the other side of the spectrum are simulations aimed at predicting Ly-$\alpha$ forest observations, 
where the dynamical range is more like $\sim$10$^4$, but as most of the contribution from the forest 
comes from gas at close to mean density, most of the computational volume must be resolved.  These
simulations require much ``shallower'' mesh refinement strategies covering a larger fraction of 
the domain. 

In addition to the challenge of achieving sufficient refinement, 
simulations need to incorporate a range of different physics beyond the obvious self-gravity. 
The presence of baryons introduces corrections to 
gravity-only predictions; these corrections are negligible at very large scales, 
but become increasingly significant at smaller scales, requiring an accurate treatment of the gas 
in order to provide realistic results that can be directly compared to observations. 
Radiative cooling and heating mechanisms are as influential on many observables 
as the addition of the gas itself, but large uncertainties remain in these terms,
and our lack of knowledge increases as we move into highly non-linear systems. 
Thus, we can model low density Lyman-$\alpha$ absorbers at higher 
redshift (e.g.~\citet{viel2005}), we can reproduce the bulk properties of 
the biggest objects in the Universe -- galaxy clusters 
(see, e.g.~\citet{stanek2010}), but on smaller and much denser scales,
galaxy formation still remains elusive, and is a field of active research (e.g.,~\citet{benson2010}). 

In this paper, we present a newly developed N-body and gas dynamics code called Nyx. The code 
models dark matter as a system of Lagrangian fluid elements, or ``particles,'' gravitationally 
coupled to an inviscid ideal fluid representing baryonic matter. 
The fluid is modeled using a finite volume representation in an Eulerian framework with block-structured AMR. 
The mesh structure used to evolve fluid quantities is also used 
to evolve the particles via a particle-mesh method. 
In order to more accurately treat hypersonic motions, where the kinetic 
energy is many orders of magnitude larger than the internal energy of the gas, we employ the dual 
energy formulation, where both the internal and total energy equations 
are solved on the grid during each time step. 

There are a number of existing finite volume, AMR codes; most commonly used in the cosmology community are
Enzo~\citep{bryan:1995}, ART (Kravtsov et al.~1997), FLASH~\citep{flash},  and RAMSES \citep{teyssier2002}.
Enzo, FLASH and Nyx are all similar in that they use block-structured AMR; 
RAMSES and ART, by contrast, refine locally on individual cells, 
resulting in very different refinement patterns and data structures.
Among the block-structured AMR codes, Enzo and FLASH enforce a strict 
parent-child relationship between patches; 
i.e., each refined patch is fully contained within a single parent patch. 
FLASH also enforces that all grid patches have the same size.
Nyx requires only that the union of fine patches be contained within the union of coarser patches 
with suitable proper nesting.  Again, this results in different refinement patterns, data structures, 
communication overhead, and scaling behavior.

Time-refinement strategies vary as well.  FLASH advances the solution at every level 
with the same time step, $\dt$, while Enzo, ART and RAMSES allow for subcycling in time.  
A standard subcycling strategy enforces that $\Delta t / \Delta x$ is constant across levels;
ART and Enzo allow greater refinement in time than space when appropriate.
Nyx maintains several different subcycling strategies.
The user can specify at run-time that Nyx use no subcycling, standard subcycling, 
a specified subcycling pattern 
or ``optimal subcycling.'' In the case of optimal subcycling, Nyx chooses, at specified
coarse time intervals during a run, the most efficient subcycling pattern across all levels based 
on an automated assessment of the computational cost of each option at that time
\citep{optimal_subcycling}.

One of the major goals behind the development of a new simulation code, in an already 
mature field such as computational cosmology, is to ensure effective utilization of existing 
and future hardware.  Future large-scale cosmological simulations will require codes capable 
of efficient scaling to hundreds of thousands of cores, and effective utilization of the 
multicore, shared memory, nature of individual nodes. One such code for gravity-only
simulations is HACC \citep{habib2009}.
The frameworks for existing hydro codes such as Enzo and FLASH are currently being
substantially re-written for current and future architectures.
Nyx, like the CASTRO code for radiation-hydrodynamics~\citep{CASTRO2010,CASTRO2011},
and the MAESTRO code
for low Mach number astrophysics~\citep{ABRZ:I,ABRZ:II,ABNZ:III,ZABNW:IV,multilevel},
is built on the BoxLib \citep{BoxLib} software framework for block-structured adaptive mesh methods, 
and as such it leverages extensive efforts to achieve the massively parallel performance that future
simulations will demand.  BoxLib currently supports a hierarchical programming approach
for multicore architectures based on both MPI and OpenMP.   Individual routines, 
such as those to evaluate heating and cooling source terms, are written in a modular fashion 
that makes them easily portable to GPUs.

In the next section we present the equations that are solved in Nyx. 
In section 3, we present an overview of the code, and in section
4 we discuss structured grid AMR and the multilevel algorithm.
Section 5 describes in more detail the parallel implementation,
efficiency and scaling results.
Section 6 contains two different cosmological tests; the first is a pure
$N$-body simulation, and the second is the Santa Barbara cluster simulation,
which incorporates both gas dynamics and dark matter.
In the final section we discuss future algorithmic developments 
and upcoming simulations using Nyx. 

\section{Basic Equations}
\label{sec:basic_equations}

\subsection{Expanding Universe}

In cosmological simulations, baryonic matter is evolved by solving the equations of 
self-gravitating gas dynamics in a coordinate system that is comoving with the expanding 
universe.  This introduces the scale factor, $a,$ into the standard hyperbolic conservation
laws for gas dynamics, where $a$ is related to the redshift, $z$, by $a = 1 / (1 + z)$, and can also serve as 
a time variable.  The evolution of $a$ in Nyx is described by the Friedmann equation 
that, for the two-component model we consider, with Hubble constant, $H_0$, and
cosmological constant, $\Omega_\Lambda$, has the form
\begin{equation}
\frac{d}{dt} \ln a = \frac{\dot{a}}{a} = 
                     H_0 \sqrt{\frac{\Omega_0}{a^3} + \Omega_\Lambda} \, \, , \label{a_equation} 
\end{equation}
where $\Omega_0$ is the total matter content of the Universe today.

\subsection{Gas Dynamics}

In the comoving frame, we relate the comoving baryonic density, $\rho_b,$ to the proper density,
$\rho_{proper}$, by $\rho_b = a^3 \rho_{proper}$, and define $\ubold$ to be the peculiar proper baryonic
velocity.  The continuity equation is then written as 
\begin{equation}
    \frac{\partial \rho_b}{\partial t} = - \frac{1}{a} \nabla \cdot
                                         (\rho_b \ubold) \ .
\end{equation}
Keeping the equations in conservation form as much as possible, we can write the
momentum equation as
\begin{eqnarray}
    \frac{\partial (a \rho_b \ubold)}{\partial t} &=&
        - \nabla \cdot (\rho_b \ubold \ubold) - \nabla p + \rho_b \gb \ ,
\end{eqnarray}
where the pressure, $p$, is related to the proper pressure, $p_{proper}$, 
by $p = a^3 p_{proper}$. Here $\gb = - \nabla \phi$ is the gravitational acceleration vector. 

We use a dual energy formulation similar to that used in Enzo~\citep{bryan:1995},
in which we evolve the internal energy as well as the total energy during each time step.  
The evolution equations for internal energy, $e$, and total energy, $E = e + \half U^2,$ can be written,
\begin{eqnarray}
    \frac{\partial (a^2 \rho_b e)}{\partial t} &=&
        - a \nabla \cdot (\rho_b \ubold e) - a p \nabla \cdot \ubold
        + a \dot{a} \left((2 - 3 (\gamma - 1) ) \rho_b e \right)
        + a \Lambda_{HC} \ , \\
    \frac{\partial (a^2 \rho_b E)}{\partial t} &=&
        - a \nabla \cdot (\rho_b \ubold E + p \ubold)
        + a \rho_b \ubold \cdot \gb
        + a \dot{a} \left((2 - 3 (\gamma - 1)) \rho_b e \right)
        + a \Lambda_{HC}
\end{eqnarray}
where $\Lambda_{HC}$ represents the combined heating and cooling terms.
We synchronize $E$ and $e$ at the end of each time step; the procedure
depends on the magnitude of $e$ relative to $E,$ and is described in the next section.

We consider the baryonic matter to satisfy a gamma-law equation of state (EOS), where we compute the
pressure, $p = (\gamma - 1) \rho e,$ with $\gamma = 5/3.$   The composition is assumed to be a mixture 
of hydrogen and helium in their primordial abundances with the different ionization states for each element
computed by assuming statistical equilibrium.
The details of the EOS and how it is related to the heating and cooling terms will be discussed in 
future work.  Given $\gamma = 5/3$, we can simplify the energy equations to the form
\begin{eqnarray}
    \frac{\partial (a^2 \rho_b E)}{\partial t} &=&
        - a \nabla \cdot (\rho_b \ubold E + p \ubold)
        + a ( \rho_b \ubold \cdot \gb + \Lambda_{HC} )  \\
    \frac{\partial (a^2 \rho_b e)}{\partial t} &=&
        - a \nabla \cdot (\rho_b \ubold e) - a p \nabla \cdot \ubold + a \Lambda_{HC} \ .
\end{eqnarray}

\subsection{Dark Matter}

Matter content in the Universe is strongly (80--90\%) dominated by cold dark matter, which 
can be modeled as a non-relativistic, pressureless fluid. The evolution of the phase-space function,
$f$, of the dark matter is thus given by the collisionless Boltzmann (aka Vlasov) equation, which 
in expanding space is: 
\begin{equation}
\label{eq:vlasov}
\frac{\partial f}{\partial t} + \frac{1}{ma^2} \bold p \cdot \nabla f 
                             - m \nabla \phi \cdot \frac{\partial f}{\partial \bold p} = 0 \, \, , 
\end{equation}
where $m$ and ${\bf p}$ are mass and momentum, respectively, and
$\phi$ is the gravitational potential. 

Rather than trying to solve Vlasov equation directly, 
we use Monte-Carlo sampling of the phase-space distribution at some initial time 
to define a particle representation, and then evolve the particles as an N-body system. 
Since particle orbits are integrals of a Hamiltonian, 
Liouville's theorem ensures that at some later time particles are still sampling 
the phase space distribution as given in equation (\ref{eq:vlasov}). 
Thus, in our simulations we represent dark matter as discrete particles with
particle $i$ having comoving location, $\xb_i,$ and peculiar proper velocity, $\ub_i,$  
and solve the system 
\begin{eqnarray}
    \frac{d \xb_i}{d t}       = \frac{1}{a} \ub_i \\
    \frac{d (a \ub_i)}{d t} = \gb_i
\end{eqnarray}
where $\gb_i$ is gravitational acceleration evaluated at the location of
particle $i$, i.e., $\gb_i = \gb(\xb_i, t)$. 

\subsection{Self-gravity}

The dark matter and baryons both contribute to the gravitational field and are
accelerated by it.  Given the comoving dark matter density, $\rho_{dm},$ we
define $\phi$ by solving 
\begin{equation}
\nabla^2 \phi (\bold{x}, t) = \frac{4 \pi G}{a} (\rho_b + \rho_{dm}- \rho_0)  \, \, , 
\label{eq:poisson}
\end{equation}
where $G$ is the gravitational constant and 
$\rho_0$ is the average value of $(\rho_{dm}+\rho_b)$ over the domain. 

\section{Code overview}

\subsection{Expanding Universe}\label{Sec:CodeExpansion}

Before evolving the gas and dark matter forward in time,
we compute the evolution of $a$ over the next full coarse
level time step.   Given $a^n$ defined as $a$ at time $t^n$,  
we calculate $a^{n+1}$ via $m$ second-order Runge-Kutta steps. 
The value of $m$ is chosen such that the relative difference between
$a^{n+1}$ computed with $m$ and $2m$ steps
is less than $10^{-8}$. This accuracy is typically achieved 
with $m=8-32$,  and the computational cost of advancing $a$ is 
trivial relative to other parts of the simulation.
We note that $\dt$ is constrained by the growth rate of $a$ such that 
$a$ changes by no more than 1\% in each time step at the coarsest level. 
Once $a^{n+1}$ is computed, values of $a$ needed at intermediate 
times are computed by linear interpolation between $a^n$ and $a^{n+1}.$

\subsection{Gas Dynamics}\label{Sec:GasDynamics}

We describe the state of the gas 
as $\Ub = (\rho_b, a \rho_b U, a^2 \rho_b E, a^2 \rho_b e),$ 
then write the evolution of the gas as
\begin{equation}
\frac{\partial\Ub}{\partial t} = -\nabla\cdot\Fb + S_e + S_g + S_{HC},
\end{equation}
where $\Fb = (1/a \; \rho_b U, \rho_b U U, a (\rho_b U E + p U), a \rho_b U e)$
is the flux vector, 
$S_e = (0, 0, 0, -a p \nabla \cdot U)$ represents the additional term in the evolution
equation for internal energy, $S_g = (0, \rho_b \gb, a \rho_b U \cdot \gb, 0)$ 
represents the gravitational source terms, 
and $S_{HC} = (0, 0, a \Lambda_{HC}, a \Lambda_{HC})$
represents the combined heating and cooling source terms.  The state, $\Ub,$ and
all source terms are defined at cell centers; the fluxes are defined on cell faces.

We compute $\Fb$ 
using an unsplit Godunov method with characteristic tracing and full
corner coupling \citep{colella1990,saltzman}.  See \cite{CASTRO2010} for a complete 
description of the algorithm without the scale factor, $a$; here we include $a$ in all
terms as appropriate in a manner that preserves second-order temporal accuracy.
The hydrodynamic integration supports both unsplit
piecewise linear \citep{colella1990,saltzman} and
unsplit piecewise parabolic (PPM) schemes \citep{ppmunsplit} to construct
the time-centered edge states used to define the fluxes.
The piecewise linear version of the implementation includes the option
to include a reference state, as discussed in \cite{ppm,colellaglaz1985}.
We choose as a reference state the average of the reconstructed profile over
the domain of dependence of the fastest characteristic 
in the cell propagating toward the interface. (The value of the reconstructed
profile at the edge is used if the fastest characteristic propagates away from the edge.)
This choice of reference state minimizes the degree to which the algorithm
relies on the linearized equations. In particular, this choice eliminates one
component of the characteristic extrapolation to edges used in second-order Godunov
algorithms.  For the hypersonic flows typical of cosmological simulations,
we have found the use of reference states improves the overall robustness of the algorithm.
The Riemann solver in Nyx iteratively solves the Riemann problem using
a two-shock approximation as described in \cite{colellaglaz1985}; 
this has been found to be more robust for hypersonic cosmological flows 
than the simpler linearized approximate Riemann solver described in \cite{CASTRO2010}.
In cells where $e$ is less than $0.01\%$ of $E,$ we use $e$ as evolved
independently to compute temperature and pressure from the EOS
and re-define $E$ as $e + \half U^2;$ otherwise we re-define $e$ as $E - \half U^2.$ 

The gravitational source terms, $S_g,$ in the momentum and energy equations are 
discretized in time using a predictor-corrector approach.  Two alternatives are available for the
discretization of $\rho_b U \cdot \gb$ in the total energy equation. 
The most obvious discretization is to compute the product of the 
cell-centered momentum, $(\rho_b U),$ and cell-centered gravitational vector, $\gb,$ 
at each time.  While this is spatially and temporally second-order accurate, it can have the
un-physical consequence of changing the internal energy, $e = E - \half U^2,$ since
the update to $E$ is analytically but not numerically equivalent to the update to the kinetic energy
calculated using the updates to the momenta.  A second alternative defines the
update to $E$ as the update to $\half U^2$. 
A more complete review of the design choices for including self-gravity in the Euler
equations is given in \cite{Springel2010}.

\subsection{Dark Matter}

We evolve the positions and velocities of the dark matter particles using a 
standard kick-drift-kick sequence, identical to that described in, for example,
\cite{miniati-colella}.  
We compute $\gb_i$ at the $i$'th particle's location, $\xb_i$ 
by linearly interpolating $\gb$ from cell centers to $\xb_i$.
To move the particles, we first accelerate the particle by $\dt/2,$
then advance the location using this time-centered velocity:
\begin{eqnarray}
    (a \ub_i)^{\nph} &=& (a \ub_i)^{n} + \frac{\dt}{2}    \gb_i^n \\
    \xb_i^{n+1}      &=& \xb_i^n + \dt \frac{1}{a^{\nph}} \ub_i^\nph 
\end{eqnarray}
After gravity has been computed at time $t^{n+1},$ we complete the
update of the particle velocities,
\begin{eqnarray}
    (a \ub_i)^{n+1} &=& (a \ub_i)^{\nph} + \frac{\dt}{2} \gb_i^{n+1} 
\end{eqnarray}
We observe that this scheme is symplectic,
thus conserves the integral of motion on average \citep{yoshida1993}.  Computationally the scheme
is also appealing because no additional storage is required during the time step 
to hold intermediate positions or velocities. 

\subsection{Self-gravity}\label{Sec:Self-gravity}

We define $\rho_{dm}$ on the mesh using the cloud-in-cell scheme, whereby we assume
the mass of the $i$'th particle is uniformly distributed over a cube of side 
$\dx$ centered at $\xb_i.$ We assign the mass of each particle
to the cells on the grid in proportion to the volume of the intersection of each 
cell with the particle's cube, and divide these cell values by $\dx^3$ 
to define the \textit{comoving} density that contributes to the right hand side
of Eq.~(\ref{eq:poisson}).

We solve for $\phi$ on the same cell centers where $\rho_b$ and $\rho_{dm}$ are defined.
The Laplace operator is discretized using the standard 7-point finite difference 
stencil and the resulting linear system is solved using
geometric multigrid techniques, specifically V-cycles with red-black Gauss-Seidel
relaxation.  The tolerance of the solver is typically set to $10^{-12}$, although
numerical exploration indicates that this tolerance can be relaxed.   
While the multigrid solvers in BoxLib support Dirichlet, Neumann and periodic 
boundary conditions, for the cosmological applications presented here we always 
assume periodic boundaries. Further gains in
efficiency result from using the solution from the previous Poisson solve as 
an initial guess for the new solve.  

Given $\phi$ at cell centers, 
we compute the average value of $\gb$ over the cell as the centered difference of $\phi$ between
adjacent cell centers.  This discretization, in combination with the interpolation of 
$\gb$ from cell centers to particle locations described above, ensures that on a uniform mesh 
the gravitational force of a particle on itself is identically zero
\citep{hockney:eastwood}.

\subsection{Computing the time step}

The time step is computed using the standard CFL condition for explicit methods,
with additional constraints from the dark matter particles and from the evolution of $a.$
Using the user-specified hyperbolic CFL factor, $0 < \sigma^\mathrm{CFL,hyp} < 1,$
and the sound speed, $c$, computed by the equation of state,
we define
\begin{equation}
\dt_{hyp} = \sigma^\mathrm{CFL,hyp} \min_{i=1\ldots 3}  \\
\left\{ \frac{a \Delta x}{ \max_{\bf x} |{U \cdot \eb_i| + c \;  }} \right \} ,
\end{equation}
where $\eb_i$ is the unit vector in the $i^{th}$ direction
and $\max_{\xb}$ is the maximum taken over all computational grid cells in the domain.

The time step constraint due to particles requires that the particles not move farther than
$\Delta x$ in a single time step.   
Since $\ub_i^{\nph}$ is not yet known when we compute the time step,
$\dt$ at time $t^n$, for the purposes of computing $\dt$ we estimate $\ub_i^{\nph}$ by $\ub_i^{n}.$
Using the user-specified CFL number for dark matter, $0 < \sigma^\mathrm{CFL,dm} < 1$, 
that is independent of the hyperbolic CFL number, we define
\begin{equation}
\dt_{dm} = \sigma^\mathrm{CFL,dm}  \min_{i=1\ldots 3}  \\
\left\{ \frac{a \Delta x}{ \max_{j} |{{\bf u}_j \cdot \eb_i| }} \right \} ,
\end{equation}
where $\max_j$ is the maximum taken over all particles $j$.  The default value for 
$\sigma^\mathrm{CFL,dm} = 0.5$.
In cases where the particle velocities are initially small but the gravitational acceleration is
large, we also constrain $\dt_{dm}$ by requiring that 
$\gb_i \dt_{dm}^2 < \dx$ for all particles to prevent
rapid acceleration resulting in a particle moving too far.

Finally, we compute $\dt_a$ as the time step over which $a$ would change by 1\%.
The time step taken by the code is set by the smallest of the three: 
\begin{equation}
\dt = \min \left( \dt_{dm}, \dt_{hyp}, \dt_{a} \right).  
\end{equation}
We note that non-zero source terms can also potentially constrain the time step; 
we defer the details of that discussion to a future paper.

\subsection{Single-Level Integration Algorithm}\label{Sec:SingleLevel}

The algorithm at a single level of refinement begins by computing the time step, $\dt,$
and advancing $a$ from $t^n$ to $t^{n+1} = t^n + \dt$. The rest of the time step is
then composed of the following steps:
\begin{description}

\item[Step 1:] {\em Compute $\phi^n$ and $\gb^n$ using $\rho_b^n$ and $\rho_{dm}^n$,
where $\rho_{dm}^{n}$ is computed from the particles at $\xb_i^{n}$.} 

We note that in the single-level algorithm we can instead use $\gb$ as computed at the 
end of the previous step because there have been no changes to $\xb_i$ 
or $\rho_b$ since then.  

\item[Step 2:] {\em Advance $\Ub$ by $\dt$.}

If we are using a predictor-corrector approach for the heating and cooling 
source terms, then we include all source terms in the explicit update:
\begin{equation}
\Ub^{n+1,\ast} = \Ub^{n} - \dt\nabla\cdot\Fb^\nph + \dt S_e^\nph + \dt S_g^n + \dt S_{HC}^n.
\end{equation}
where $\Fb^\nph$ and $S_e^\nph$ are computed by predicting from the $\Ub^{n}$ states.

If we are instead using Strang splitting for the heating and cooling source terms, 
we first advance $(\rho e)$ and $\rho E$ 
by integrating the source terms in time for $\half \dt$ 
\begin{eqnarray}
(\rho e)^{n,\ast} &=& (\rho e)^n +  \int \Lambda_{HC} \; dt^\prime \enskip , \\
(\rho E)^{n,\ast} &=& (\rho E)^n +  \int \Lambda_{HC} \; dt^\prime \enskip ,
\end{eqnarray}
then advance the solution using time-centered fluxes and $S_e$ 
and an explicit representation of $S_g$ at time $t^n$:  
\begin{equation}
\Ub^{n+1,\ast} = \Ub^{n,\ast} - \dt\nabla\cdot\Fb^\nph + \dt S_e^\nph + \dt S_g^n
\end{equation}
where $\Fb^\nph$ and $S_e^\nph$ are computed by predicting from the $\Ub^{n,\ast}$ states.
The details of how the heating and cooling source terms are incorporated depend 
on the specifics of the heating and cooling mechanisms represented by $\Lambda_{HC}$; 
we defer the details of that discussion to a future paper.

\item[Step 3:] {\em Interpolate $\gb^n$ from the grid to the particle locations, then
advance the particle velocities by $\Dt / 2$ and particle positions by $\Dt$.}
\begin{eqnarray}
     \ub_i^{\nph} &=& \frac{1}{a^{\nph}} ((a^n \ub^n_i) + \frac{\dt}{2} \; \gb^n_i) \\
     \xb_i^{n+1}  &=& \xb^n_i + \frac{\dt}{a^{\nph}} \ub_i^{\nph}
\end{eqnarray}

\item[Step 4:] {\em Compute $\phi^{n+1}$ and $\gb^{n+1}$ using 
$\rho_b^{n+1,*}$ and $\rho_{dm}^{n+1}$, where $\rho_{dm}^{n+1}$
is computed from the particles at $\xb_i^{n+1}$.} 

Here we can use $\phi^n$ as an initial guess for $\phi^{n+1}$ in order to reduce the time
spent in multigrid to reach the specified tolerance.

\item[Step 5:] {\em Interpolate $\gb^{n+1}$ from the grid to the particle locations, then
update the particle velocities, $\ub_i^{n+1}$}
\begin{eqnarray}
    \ub_i^{n+1} &=& \frac{1}{a^{n+1}}
                    \left( \left( a^{\nph} \ub^{\nph}_i \right)
                         + \frac{\dt}{2} \; \gb^{n+1}_i \right)  
\end{eqnarray}

\item[Step 6:] {\em Correct $\Ub$ with time-centered source terms, and replace $e$ by
$E - \half U^2$ as appropriate.}

If we are using a predictor-corrector approach for the heating and cooling 
source terms, we correct the solution by effectively time-centering all the source terms:
\begin{equation}
\Ub^{n+1} = \Ub^{n+1,\ast} + \frac{\dt}{2} (S_g^{n+1,\ast} - S_g^n) + 
                             \frac{\dt}{2} (S_{HC}^{n+1,\ast} - S_{HC}^n) \enskip .
\end{equation}

If we are using Strang splitting for the heating and cooling source terms, we time-center the 
gravitational source terms only,
\begin{equation}
\Ub^{n+1,\ast \ast} = \Ub^{n+1,\ast} + \frac{\dt}{2} (S_g^{n+1,\ast} - S_g^n)
\end{equation}
then integrate the source terms for another $\half \dt,$ 
\begin{eqnarray}
(\rho e)^{n+1} &=& (\rho e)^{n+1,\ast \ast} + \int \Lambda_{HC} \; dt^\prime \enskip , \\
(\rho E)^{n+1} &=& (\rho E)^{n+1,\ast \ast} + \int \Lambda_{HC} \; dt^\prime \enskip .
\end{eqnarray}
We note here that the time discretization of the gravitational source terms differs from that 
in Enzo \citep{bryan:1995}, where $S_g^{\nph}$ is computed by 
extrapolation from values at $t^n$ and $t^{n-1}$.   

If, at the end of the time step, 
$(\rho E)^{n+1} - \half \rho^{n+1} (U^{n+1})^2 > 10^{-4} (\rho E)^{n+1},$ 
we re-define $(\rho e)^{n+1} = (\rho E)^{n+1}  - \half \rho^{n+1} (U^{n+1})^2$;
otherwise we use $e^{n+1}$ as evolved above when needed to compute pressure or temperature,
and re-define $(\rho E)^{n+1} = (\rho e)^{n+1} + \half \rho^{n+1} (U^{n+1})^2.$

\end{description}

\noindent This concludes the single-level algorithm description.  

\section{AMR}\label{Sec:AMR}
Block-structured AMR was introduced by \cite{berger-oliger};
it has been demonstrated to be highly successful for gas dynamics by 
\cite{berger-colella} in two dimensions and by \cite{bell-3d} in three dimensions,
and has a long history of successful use in a variety of fields.
The AMR methodology in Nyx uses a nested hierarchy of rectangular grids 
with refinement of the grids in space by a factor of two between levels, 
and refinement in time between levels as dictated by the specified subcycling algorithm.

\subsection{Mesh Hierarchy}

The grid hierarchy in Nyx is composed of different levels of refinement ranging
from coarsest ($\ell = 0$) to finest ($\ell = {\ell}_{\rm finest}$).
The maximum number of levels of refinement allowed, $\ell_{\rm max}$, is specified 
at the start (or any restart) of a calculation. At any given time in the calculation 
there may be fewer than $\ell_{\rm max}$ levels in the hierarchy, i.e.\ $\ell_{\rm finest}$
can change dynamically as the calculation proceeds as long as 
$\ell_{\rm finest} \leq \ell_{\rm max}.$ 
Each level is represented by the union of non-overlapping rectangular grids 
of a given resolution.  
Each grid is composed of an even
number of ``valid'' cells in each coordinate direction; cells 
are the same size in each coordinate direction but grids
may have different numbers of cells in each direction.  
Each grid also has ``ghost cells'' which extend outside the grid by the same 
number of cells in each coordinate direction on both the low and high ends
of each grid.  These cells are used to temporarily hold data used to
update values in the ``valid'' cells; when subcycling in time the ghost cell
region must be large enough that a particle at level $\ell$ can not leave the
union of level $\ell$ valid and ghost cells over the course of a level $\ell-1$ timestep.
The grids are properly nested, in the sense that the union of grids
at level $\ell+1$ is contained in the union of grids at level $\ell$.
Furthermore, the containment is strict in the sense that the level 
$\ell$ grids are large enough to guarantee a border at least 
$n_{\rm proper}$ level $\ell$ cells wide surrounding each level
$\ell +1$ grid, where $n_{\rm proper}$ is specified by the user. 
There is no strict parent-child relationship; a single level $\ell+1$ grid 
can overlay portions of multiple level $\ell$ grids.

We initialize the grid hierarchy and regrid following the procedure 
outlined in \cite{bell-3d}.  To define grids at level $\ell+1$
we first tag cells at level $\ell$ where user-specified criteria are met.
Typical tagging criteria, such as local overdensity, can be selected
at run-time; specialized refinement criteria,
which can be based on any variable or combination of variables
that can be constructed from the fluid state or particle information,
can also be used.
The tagged cells are grouped into rectangular grids at level $\ell$ using
the clustering algorithm given in \cite{bergerRigoutsos:1991}.
These rectangular patches are then refined to form the grids at level $\ell+1$.
Large patches are broken into smaller patches for distribution to multiple 
processors  based on a user-specified
{\it max\_grid\_size} parameter.

A particle, $i,$ is said to ``live'' at level $\ell$ if level $\ell$ is the finest level 
for which the union of grids at that level contains the particle's location, $\xb_i$. 
The particle is said to live in the $n^{th}$ grid at level $\ell$ if the particle lives
at level $\ell$ and the $n^{th}$ grid at level $\ell$ contains $\xb_i.$
Each particle stores, along with its location, mass and velocity, the current information
about where it lives, specifically the level, grid number, and cell, $(i,j,k)$.  Whenever
particles move, this information is recomputed for each particle in a ``redistribution'' procedure. 
The size of a particle in the PM scheme
is set to be the mesh spacing of the level at which the particle currently lives.

At the beginning of every $k_\ell$ level $\ell$ time steps,
where $k_\ell \geq 1$ is specified by the user at run-time,
new grid patches are defined at all levels $\ell+1$ and higher
if $\ell < \ell_{\rm max}.$   In regions previously covered by 
fine grids the data is simply copied from old grids to new; in 
regions that are newly refined, data is interpolated from underlying
coarser grids. The interpolation procedure constructs 
piecewise linear profiles within each coarse cell based on nearest neighbors; 
these profiles are limited so as to not introduce any new maxima or minima, 
then the fine grid values are defined to be the value of the trilinear profile 
at the center of the fine grid cell,  thus ensuring conservation of all interpolated quantities.
All particles at levels $\ell$ through $\ell_{max}$ are redistributed 
whenever the grid hierarchy is changed at levels $\ell+1$ and higher.

\subsection{Subcycling Options}

Nyx supports four different options for how to relate $\dt_\ell,$ the time step
at level $\ell > 0,$ to $\dt_0$, the time step at the coarsest level.  
The first is a non-subcycling option; in this case all levels are advanced with the same $\dt,$
i.e. $\dt_\ell = \dt_0$ for all $\ell.$
The second is a ``standard'' subcycling option, in which $\dt / \dx$ is constant across all levels,
i.e. $\dt_\ell = r^\ell \dt_0$ for spatial refinement ratio, $r$, between levels.
These are both standard options in multilevel algorithms.  In the third case, the user
specifies the subcycling pattern at run-time;
for example, the user could specify subcycling between levels 0 and 1, but no subcycling 
between levels 1, 2 and 3.  Then $\dt_3 = \dt_2 = \dt_1 = r \; \dt_0.$
The final option is ``optimal subcycling,'' in which the optimal subcycling algorithm decides
at each coarse time step, or other specified interval, what the subcycling pattern should be.
For example, if the time step at each level is constrained by $\dt_a,$ 
the limit that enforces the condition that $a$ not change more than $1 \%$ in a time step,
then not subcycling is the most efficient way to advance the solution on the multilevel hierarchy.  
If dark matter particles at level 2 are moving a factor of two
faster than particles at level 3 and $\dt_{dm}$ is the limiting constraint in
choosing the time step at those levels, it may be most efficient to advance levels 2 and 3 
with the same $\dt$, but to subcycle level 1 relative to level 0 and level 2 relative to level 1.
These choices are made by the optimal subcycling algorithm during the run;
the algorithm used to compute the optimal subcycling pattern is described 
in \cite{optimal_subcycling}.

In addition to the above examples, additional considerations,
such as the ability to parallelize over levels rather than just over grids at a particular level, 
may arise in large-scale parallel computations, which can make more complex subcycling 
patterns more efficient than the ``standard'' pattern. We defer further discussion of these
cases to future work in which we demonstrate the trade-offs in computational efficiency for
specific cosmological examples.

\subsection{Multilevel Algorithm}

We define a complete multilevel ``advance'' as the combination of operations 
required to advance the solution at level 0 by one level 0 time step
and operations required to advance the solution at all finer
levels, $0 < \ell \le \ell_{finest},$
to the same time.  
We define $n_\ell$ as the number of time steps taken at level $\ell$ for each 
time step taken at the coarsest level, i.e.    
$\dt_0 = n_\ell \dt_\ell$ for all levels, $0 < \ell \le \ell_{finest}.$ 
If the user specifies no subcycling, then 
$n_\ell = 1$; if the user specifies standard subcycling then
$n_\ell = r^\ell.$    If the user defines a subcycling pattern, this
is used to compute $n_\ell$ at the start of the computation;  if optimal subcycling is
used then $n_\ell$ is re-computed at specified intervals.

The multilevel advance begins by computing $\dt_\ell$
for all $\ell \ge 0,$ which first requires the computation of 
the maximum possible time step $\dt_{\max,\ell}$ at each level independently 
(including only the particles at that level).  
Given a subcycling pattern and $\dt_{\max,\ell}$ at each level,
the time steps are computed as: 
\begin{equation}
\dt_0 = \min_{0 \le \ell \le \ell_{finest}}(n_\ell \cdot \dt_{\max,\ell})
\end{equation}
\begin{equation}
\dt_\ell = \frac{\dt_0}{n_\ell}
\end{equation}

Once we have computed $\dt_\ell$ for all $\ell$
and advanced $a$ from time $t$ to $t + \dt_0,$
a complete multilevel ``advance'' is defined by calling the
recursive function, {\bf Advance}($\ell$), described below, with $\ell = 0.$
We note that in the case of no-subcycling, the entire multilevel advance
occurs with a single call to {\bf Advance}(0); in the case of standard subcycling
{\bf Advance}($\ell$) will be called recursively for each $\ell,$ $0 \le \ell \le \ell_{finest}$.  
Intermediate subcycling patterns require calls to {\bf Advance}($\ell$) for every level $\ell$ that
subcycles relative to level $\ell-1.$  In the notation below, the superscript $n$ refers
to the time, $t^n,$ at which each call to {\bf Advance} begins at level $\ell$, 
and the superscript $n+1$ refers to time $t^n + \dt_\ell.$

\noindent{\bf Advance ($\ell$):}
\begin{description}

\item[Step 0:] {\em Define $\ell_a$ as the finest level such that 
$\dt_{\ell_a} = \dt_\ell.$}

If we subcycle between levels $\ell$ and $\ell+1$ then $\ell_a = \ell$
and this call to {\bf Advance} only advances level $\ell$. If, however, 
we don't subcycle we can use a more efficient multi-level advance on 
levels $\ell$ through $\ell_a.$

\item[Step 1:] {\em Compute $\phi^n$ and $\gb^n$ at levels 
$\ell\to \ell_{finest}$ 
using $\rho_b^n$ and $\rho_{dm}^n$,
where $\rho_{dm}^{n}$ is computed from the particles at $\xb_i^{n}$.}

This calculation involves solving the Poisson equation on the 
multilevel grid hierarchy using \emph{all} finer levels; 
further detail about computing $\rho_{dm}$ and
solving on the multilevel hierarchy is given in the next section.
If $\phi$ has already been computed at this time during a multilevel solve in
{\bf {Step 1}} at a coarser level, this step is skipped.

\item[Step 2:] {\em Advance $\Ub$ by $\dt_\ell$ at levels $\ell\to \ell_a$.}

Because we use an explicit method to advance the gas dynamics, each level is
advanced independently by $\dt_\ell$, using Dirichlet boundary conditions 
from a coarser level as needed for boundary conditions.  
If $\ell_a > \ell,$ then after levels $\ell \to \ell_a$ have been advanced, 
we perform an explicit reflux of the hydrodynamic quantities
(see, e.g., \citet{CASTRO2010} for more details on refluxing) from level
$\ell_a$ down to level $\ell$ to ensure conservation. 

\item[Step 3:] {\em Interpolate $\gb^n$ at levels $\ell \to \ell_a$
from the grid to the particle locations, then 
advance the particle velocities at levels $\ell\to 
\ell_a$ by $\Dt_\ell / 2$ and particle positions by $\Dt_\ell$.}
The operations in this step are identical to those in the 
single-level version, and require no communication between grids or between levels.
Although a particle may move from a valid cell into a ghost cell, or 
from a ghost cell into another ghost cell, during this step,
it remains a level $\ell$ particle until redistribution
occurs after the completion of the level $\ell-1$ timestep.

\item[Step 4:] {\em Compute $\phi^{n+1}$ and $\gb^{n+1}$ at levels $\ell\to 
\ell_a$ using $\rho_b^{n+1,\ast}$ and $\rho_{dm}^{n+1}$,
where $\rho_{dm}^{n+1}$ is computed from the particles at $\xb_i^{n+1}$.}

Unlike the solve in {\bf Step 1}, this solve over levels $\ell\to \ell_a$ 
is only a partial multilevel solve unless this is a no-subcycling algorithm;  
further detail about the multilevel solve is given in the next section.

\item[Step 5:] {\em Interpolate $\gb^{n+1}$ at levels $\ell \to \ell_a$ 
from the grid to the particle locations, then
update the particle velocities at levels $\ell\to \ell_a$
by an additional $\dt_\ell / 2$}.
Again this step is identical to the single-level algorithm.

\item[Step 6:] {\em Correct $\Ub$ at levels $\ell\to \ell_a$
with time-centered source terms, and replace $e$ by $E - \half U^2$ as appropriate.}

\item[Step 7:] {\em If} $\ell_a < \ell_{finest},$ {\bf Advance}($\ell_a+1$) 
$n_{\ell_a+1} / n_\ell$ {\em times}.

\item[Step 8:] {\em Perform any final synchronizations}

\begin{itemize}
\item If $\ell_a < \ell_{finest},$ reflux the hydrodynamic quantities 
from level $\ell_a+1$ to level $\ell_a.$   
\item If $\ell_a < \ell_{finest},$ average down the solution from level $\ell_a+1$ to level $\ell_a$. 
\item If $\ell   < \ell_a       ,$ average down the solution from level $\ell_a$ through level $\ell$. 
\item If $\ell_a < \ell_{finest},$ perform an elliptic synchronization solve to compute $\delta \phi$
at levels $\ell$ through $\ell_{finest},$
where the change in $\phi$ results from the change in $\rho$ at level $\ell_a$ 
due to refluxing from level $\ell_a+1$, and from the mismatch in Neumann boundary conditions
at the $\ell_a$ / $\ell_a+1$ interface (that results because the finest level in the
solve from {\bf Step 4} was $\ell_a,$ not $\ell_{finest}$). Add $\delta \phi$ to $\phi$
at levels $\ell$ through $\ell_{finest}.$  This synchronization solve is described in
more detail in \citet{CASTRO2010} for the case of standard subcycling.
\item Redistribute particles at levels $\ell\to \ell_{finest};$ 
Note that, if $\ell > 0,$ 
particles that started the time step at level $\ell$ but now would live at level $\ell-1$
are kept for now in the ghost cells at level $\ell$; they are not redistributed to
level $\ell-1$ until the level $\ell-1$ timestep is complete.

\end{itemize}

\end{description}

\subsection{Gravity Solves}

In the algorithm as described above, there are times when the Poisson equation for 
the gravitational potential,
\begin{equation}
\nabla^2 \phi = \frac{4 \pi G}{a} (\rho_b + \rho_{dm}- \rho_0)  \, \, , 
\end{equation}
is solved simultaneously on multiple levels, and times when the solution is only 
required on a single level.  We define $L^\ell$ as the standard 7-point finite difference
approximation to $\nabla^2$ at level $\ell,$ modified as appropriate at the $\ell / (\ell-1)$ interface
if $\ell > 0;$  see, e.g., \cite{almgren-iamr} for details of the 
multilevel cell-centered interface stencil.  

When $\ell = \ell_{finest}$ in {\bf Step 1}, or $\ell = \ell_a$ in {\bf Step 4},
we solve
\[ L^{\ell} \phi^{\ell} =  \frac{4 \pi G}{a} (\rho_b^\ell + \rho_{dm}^\ell - \rho_0)  \, \,   \]
on the union of grids at level $\ell$ only; this is referred to as a {\it level solve}.

If $\ell=0$ then the grids at this level cover the entire domain and the only boundary conditions
required are periodic boundary conditions at the domain boundaries.  If $\ell > 0$ then
Dirichlet boundary conditions for the level solve at level $\ell$ are supplied at 
the $\ell / (\ell-1)$ interface from data at level $\ell-1.$   
Values for $\phi$ at level $\ell-1$ are assumed to live at the 
cell centers of level $\ell-1$ cells; these values are interpolated tangentially at the 
$\ell / (\ell-1)$ interface to supply boundary values with spacing $\dx^\ell$ for the level $\ell$ grids.  
The modified stencil obviates the need for interpolation normal to the interface.
The resulting linear system is solved using
geometric multigrid techniques, specifically V-cycles with red-black Gauss-Seidel relaxation.  
We note that after a level $\ell$ solve with $\ell > 0$, $\phi$ at levels $\ell$ and
$\ell-1$ satisfy Dirichlet but not Neumann matching conditions.  This mismatch is corrected
in the elliptic synchronization described in {\bf Step 8}.

When a multilevel solve is required,
we define $L_{\ell,m}^{\rm comp}$ as the composite grid approximation
to $\nabla^2$ on levels $\ell$ through $m$, and define a {\it composite solve}
as the process of solving
\[ L_{\ell,m}^{\rm comp} \phi^{\rm comp} =  
\frac{4 \pi G}{a} (\rho_b^{\rm comp} + \rho_{dm}^{\rm comp} - \rho_0)  \, \,   \]
on levels $\ell$ through $m.$
The solution to the composite solve satisfies
\[ L^{m} \phi^{\rm comp} =  \frac{4 \pi G}{a} (\rho_b^m + \rho_{dm}^m - \rho_0)  \, \,   \]
at level $m$, but satisfies
\[ L^{\ellp} \phi^{\ellp} =  \frac{4 \pi G}{a} (\rho_b^\ellp + \rho_{dm}^\ellp - \rho_0)  \, \,   \]
for $\ell \leq \ellp < m$ only on the regions of each level
{\it not} covered by finer grids or adjacent to the boundary
of the finer grid region.  In regions of a level $\ellp$ grid
covered by level $\ellp+1$ grids
$\phi^\ellp$ is defined as the volume average of $\phi^{\ellp+1}$;
in level $\ellp$ cells immediately adjacent to the boundary
of the union of level $\ellp+1$ grids, a modified interface operator 
as described in \cite{almgren-iamr} is used.  This linear system is also solved 
using geometric multigrid with V-cycles; here levels $\ell$ through $m$ serve as 
the finest levels in the V-cycle.  
After a composite solve on levels $\ell$ through $m$, $\phi$ satisfies
both Dirichlet and Neumann matching conditions at the interfaces between
grids at levels $\ell$ through $m$; however, there is still
a mismatch in Neumann matching conditions between $\phi$ at levels $\ell$ and $\ell-1$ if
$\ell > 0.$  This mismatch is corrected in the elliptic synchronization described in {\bf Step 8}.

\subsection{Cloud-in-Cell Scheme}

Nyx uses the cloud-in-cell scheme to define the contribution of each dark matter particle's mass
to $\rho_{dm};$ a particle contributes to $\rho_{dm}$ only on the cells that intersect 
the cube of side $\dx$ centered on the particle.  In the single-level algorithm, all particles
live on a single level and contribute to $\rho_{dm}$ at that level.  If there are multiple
grids at a level, and a particle lives in a cell adjacent to a boundary with another grid, 
some of the particle's contribution will be to the grid in which it lives, and the rest
will be to the adjacent grid(s).  
The contribution of the particle to cells in adjacent grid(s) is computed in two stages: 
the contribution is first added to a ghost cell of the grid in which the particle lives, 
then the value in the ghost cell is added to the value in the valid cell of the adjacent grid.
The right-hand-side of the Poisson solve sees only the values of $\rho_{dm}$ on the valid
cells of each grid.

In the multilevel algorithm, the cloud-in-cell procedure becomes more complicated.  During
a level $\ell$ level solve or a multilevel solve at levels $\ell \to m,$ a particle at level $\ell$
that lives near the $\ell / (\ell-1)$ interface may ``lose'' part of its mass across the interface.
In addition, during a level solve at level $\ell,$ or a multi-level solve at levels $\ell \to m,$
it is essential to include the contribution from dark matter particle at other levels even if 
these particles do not lie near a coarse-fine boundary.  Both of these complications are
addressed by the creation and use of {\it ghost} and {\it virtual} particles. 

The mass from most particles at levels $\ell^\prime < \ell$ 
is accounted for in the Dirichlet boundary conditions for $\phi^\ell$ that are supplied from 
$\phi^{\ell-1}$ at the $\ell / (\ell-1)$ interface.  However, level $\ell-1$ particles that live
near enough to the $\ell / (\ell-1)$ interface to deposit part of their mass at level $\ell,$
or have actually moved from level $\ell-1$ to level $\ell$ during a level $\ell-1$ timestep,
are not correctly accounted for in the boundary conditions. 
The effect of these is included via ghost particles, which are copies at level 
$\ell$ of level $\ell-1$ particles.  Although these ghost particles
live and move at level $\ell,$ they retain the size, $\dx^{\ell-1}$ of the particles of which they 
are copies.

The mass from particles living at levels 
$\ell^\prime > m$ in a level $m$ level solve or a level $\ell \to m$ multilevel solve,
is included at level $m$ via the creation of  virtual particles.  These are
level $m$ representations of particles that live at levels $\ell^\prime > m.$
Within Nyx there is the option to have one-to-one correspondence of fine level particles
to level $m$ virtual particles; the option also exists to aggregate these fine level
particles into fewer, more massive virtual particles at level $m.$  The total mass 
of the finer level particles is conserved in either representation.  These virtual particles
are created at level $m$ at the beginning of a level $m$ timestep and moved along
with the real level $m$ particles; this is necessary because the level $\ell^\prime$
particles will not be moved until after the second Poisson solve with top level $m$ 
has occurred.  The mass of the virtual particles contributes to $\rho_{dm}$ at level $m$ 
following the same cloud-in-cell procedure as followed by the level $m$ particles,
except that the virtual particles have size $\dx^{m+1}$ (even if they are copies of
particles at level $\ell^\prime$ with $\ell^\prime > m+1$).


\section{Software Design and Parallel Performance}

\subsection{Overview}

Nyx is implemented within the BoxLib \citep{BoxLib} framework,
a hybrid \Cplusplus \slash Fortran90 software system that provides
support for the development of parallel block-structured AMR applications.
The basic parallelization strategy uses a hierarchical programming approach
for multicore architectures based on both MPI and OpenMP.   
In the pure-MPI instantiation, at least one grid at each level is 
distributed to each core, and each core communicates
with every other core using only MPI.  In the hybrid approach, where 
on each socket there are $n$ cores that all access the same memory, 
we can instead have fewer, larger grids per socket,
with the work associated with those grids distributed among the $n$ cores
using OpenMP.    

In BoxLib, memory management, flow control, parallel communications and I/O are
expressed in the \Cplusplus portions of the program.  The numerically intensive portions of the 
computation, including the multigrid solvers, are handled in Fortran90.
The software supports two data distribution schemes for data at a level, 
as well as a dynamic switching scheme that decides which approach to use 
based on the number of grids at a level and the number of processors.
The first scheme is based on a heuristic knapsack algorithm as
described in \citet{crutchfield:1991} and in \citet{rendleman-hyper}. 
The second is based on the use of a Morton-ordering space-filling curve.

\subsection{Particle and Particle-Mesh Operations}

BoxLib also contains support for parallel particle and particle-mesh operations.
Particles are distributed to processors or nodes according to their location; 
the information associated with particle $i$ at location $\xb_i$ exists 
only on the node that owns the finest-level grid that contains $\xb_i.$
Particle operations are performed with the help of particle iterators that loop
over particles; there is one iterator associated with each MPI process, and each
iterator loops over all particles that are associated with that process. 

Thus interactions between particles and grid data, such as the calculation of
$\rho_{dm},$ or the interpolation of ${\bf g}$ from the grid to $\xb_i,$
require no communication between processors or nodes; they
scale linearly with the number of particles, and exhibit almost perfect 
weak scaling.   Similarly, the update of the particle velocities using the
interpolated gravitational acceleration, and the update of the particle locations
using the particle velocities, require no communication.
The only inter-node communication of particle data 
(location, mass and velocity) occurs when a particle
moves across coarse-fine or fine-fine grid boundaries; in this case 
the redistribution process re-computes the particle level, grid number, and cell,
and the particle information is sent to the node or core where the particle now lives.
We note, though, that when information from the particle is represented on the
grid hierarchy, such as when the particle ``deposits" its mass on the 
grid structure in the form of a density field used in the Poisson solve,
that grid-based data may need to be transferred between nodes if
the particle is near the boundary of the grid it resides in, and 
deposits part of its mass into a different grid at the same or a different level.

Looking ahead, we also note here that the particle data structures in BoxLib are 
written in a sufficiently general form with \Cplusplus templates that they
can easily be used to represent other types of particles (such as those
associated star formation mechanisms) which carry different
numbers of attributes beyond location, mass and velocity.

\subsection{Parallel I/O and Visualization}

As simulations grow increasingly large, the need to read and write
large data-sets efficiently becomes increasingly critical.
Data for checkpoints and analysis are written by Nyx in a self-describing
format that consists of a directory for each time step written.
Checkpoint directories contain all necessary data to restart the
calculation from that time step.  Plotfile directories contain data
for postprocessing, visualization, and analytics, which can be read
using \Code{amrvis}, a customized visualization package developed at
LBNL for visualizing data on AMR grids, VisIt \citep{visit}, or yt \citep{yt}.  
Within each checkpoint or plotfile
directory is an ASCII header file and subdirectories for each AMR
level.  The header describes the AMR hierarchy, including the number of
levels, the grid boxes at each level, the problem size, refinement
ratio between levels, step time, etc.  Within each level directory are
multiple files for the data at that level.  Checkpoint and plotfile
directories are written at user-specified intervals.

For output, a specific number of data files per level is specified at run time.
Data files typically contain data from multiple processors, so each
processor writes data from its associated grid(s) to one file, then 
another processor can write data from its associated grid(s) to that 
file.  A designated I/O processor writes the header files
and coordinates which processors are allowed to write to which files
and when.  The only communication between processors is for signaling
when processors can start writing and for the exchange of header
information.  
While I/O performance even during a single run can
be erratic, recent timings of parallel I/O on the Hopper machine at NERSC,
which has a theoretical peak of 35GB/s,
indicate that Nyx's I/O performance, when run with a single level 
composed of multiple uniformly-sized grids, can reach over 34GB/s.

Restarting a calculation can present some difficult issues
for reading data efficiently.  
Since the number of files is generally not equal to the
number of processors, input during restart is coordinated to
efficiently read the data.  Each data file is only opened by one
processor at a time.
The designated I/O processor creates a database for mapping files
to processors, coordinates the read queues, and interleaves
reading its own data.  Each processor reads all data it needs
from the file it currently has open.
The code tries to maintain the number of input streams to be
equal to the number of files at all times.

Checkpoint and plotfiles are portable to machines with a
different byte ordering and precision from the machine that wrote the files.
Byte order and precision translations are done automatically,
if required, when the data is read.

\subsection{Parallel Performance}

\begin{figure}[t]
\begin{center}
    \includegraphics[width=150mm]{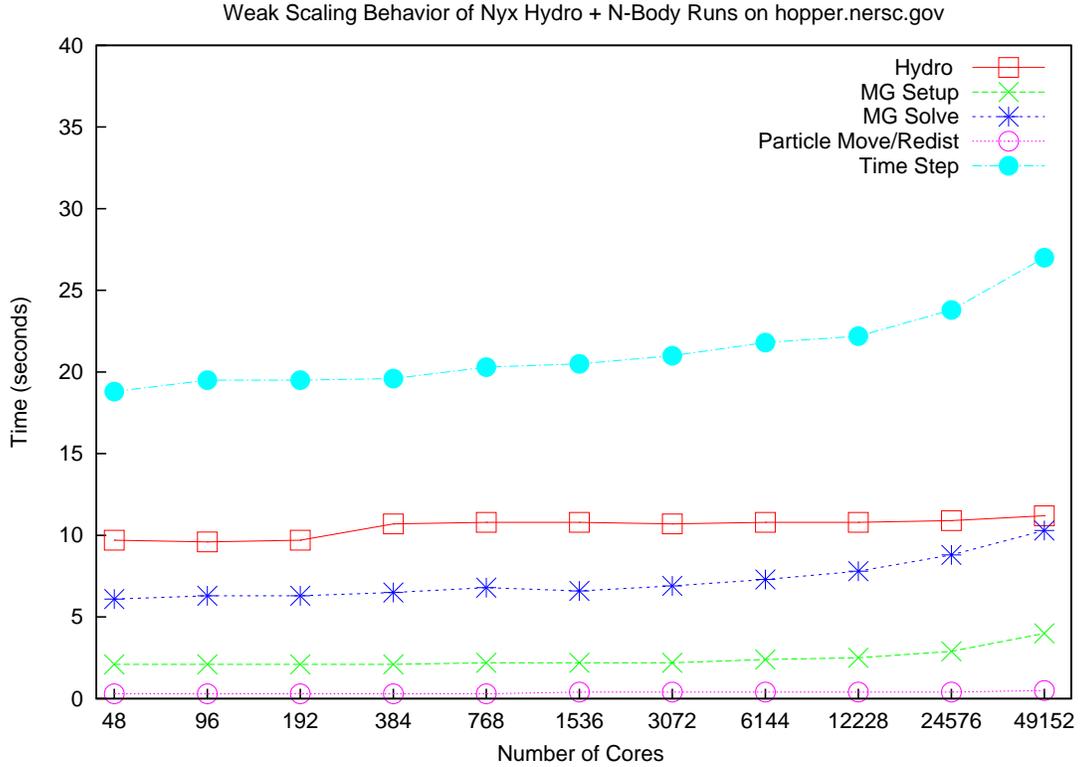} 
\end{center}
    \caption{Weak scaling behavior of Nyx on the NERSC Hopper (Cray XE 6) machine 
             for simulations of the replicated Santa Barbara problem.}
    \label{fig:scaling}
    \vspace{0.20 truecm}
\end{figure}

In Figure~\ref{fig:scaling} we show the scaling behavior of the Nyx code
using MPI and OpenMP on the Hopper machine at NERSC.  A weak scaling study 
was performed, so that for each run there was exactly one $128^3$ grid per NUMA node, 
where each NUMA node has 6 cores.\footnote{We note that for small to moderate
numbers of cores, pure-MPI runs can be significantly faster than runs 
using hybrid MPI and OpenMP.  However, for the purposes of this scaling study
we report only hybrid results.}

The problem was initialized using replication of the Santa Barbara problem, 
described in the next section.  First, the original $256^3$ particles were read in, 
then these particles were replicated in each coordinate
direction according to the size of the domain.  For example, on a run with 1024 x 1024 x 2048
cells, the domain was replicated 4 times in the x- and y-directions and 8 times in the
z-direction.   With one $128^3$ grid per NUMA node, this run would have used
1024 processors.  The physical domain was scaled with the index space, so that the
resolution ($\Delta x$) of the problem didn't change with weak scaling, thus the 
characteristics of the problem which might impact the number of multigrid V-cycles, for example,
were unchanged as the problem got larger.

The timings in Figure~\ref{fig:scaling} are the time per time step spent in different parts of
the algorithm for a uniform grid calculation.   The "hydro" timing represented all time spent
to advance the hydrodynamic state, excluding the calculation of the gravity.  The computation
of gravity is broken into two parts; the set-up and initialization of the multigrid solvers,
and the actual multigrid V-cycles themselves.   The set-up phase includes the assignment of
mass from the particles to the mesh using the cloud-in-cell scheme.  In this run, 
each multigrid solve took 7 V-cycles to reach a tolerance of $10^{-12}.$  We note that,
for this problem which has an average of one particle per cell,
the contribution to the total time from moving and redistributing the particles is negligible.

We see here that the Nyx code scales well from 48 to 49,152 processors,   
with a less than 50\% increase in total time with the 1000-fold increase in processors.
The hydrodynamic core of
CASTRO, which is essentially the same as that of Nyx, has shown 
almost flat scaling to 211K processors on the Jaguarpf machine at OLCF \citep{BoxLib},
and only a modest overhead from using AMR with subcycling, ranging from roughly 5\% for 8 processors to 19\% 
for 64K processors \citep{CASTRO2010}.  MAESTRO, another BoxLib-based code which uses the 
same linear solver framework as Nyx, has demonstrated excellent scaling to 96K cores on Jaguarpf \citep{BoxLib}.

\section{Validation}

The hydrodynamic integrator in Nyx was built by extending the integrator
in CASTRO to the equations for an expanding universe;  the Nyx hydrodynamics
have been validated using the same tests, and with the same results, 
as described in \cite{CASTRO2010}.
To isolate and study the behavior of the component of the algorithm responsible for
updating the particle locations and velocities, we have performed a resolution study of a 
simple two-particle orbit.
In a nondimensionalized cubic domain 16 units on a side, we initialized two particles of equal 
mass each at a distance of 1 unit in the $x$-direction from the center of the domain.  
In order to compare the numerical results to the exact circular orbit solution, 
the initial particle velocities were specified to be those corresponding to a perfectly circular orbit,
and Dirichlet boundary conditions, constructed as the sum of two monopole expansions (one from each particle), 
were imposed at the domain boundaries. 
Six numerical tests were performed: three with a base grid of $64^3$, and either 0, 1 or 2
levels of refinement; one with a base grid of $128^3$ and 0 or 1 level of refinement, and finally a 
uniform $256^3$ case.
Three key observations followed.  First, the orbits of the particles were stable over the ten orbits observed,
with little variation in the measured properties over time.  Second, the orbit radius and 
kinetic energy of each particle converged with mesh spacing; the maximum deviations 
of the orbit radius and the particle's kinetic energy from their initial values
were no more than approximately $1.1\%$ and $2.4\%,$ respectively, for the coarsest case 
and $0.17 \%$ and $0.36 \%,$ respectively, for the finest cases.  
Third, the difference in the deviation of the orbit radius between 
the single-level and multilevel simulations with the same finest resolution 
was less than $.0002 \%$ of the radius; the difference in deviation of the kinetic energy
was less than $.0004 \%$ of the initial kinetic energy.

We have also conducted several simple tests with known analytical solutions,
e.g. the MacLaurin spheroid and Zel'dovich pancakes, to additionally validate the particle dynamics 
and the expected second-order accuracy and convergence properties of the Nyx gravity solver. 
Finally in this section, we present two validation studies in a cosmological context, 
using well-established data sets from the Cosmic Data ArXiv \citep{heitmann05} and the
Santa Barbara Cluster Comparison Project \citep{frenk1999}. 

\subsection{Dark Matter Only Simulations}

To validate Nyx in a realistic cosmological setting, we first compare 
Nyx gravity-only simulation results to one of the comparison data sets described in
\citet{heitmann05}.  The domain is 256 Mpc/h on a side,  and
the cosmology considered is $\Omega_m = 0.314$, $\Omega_{\Lambda} = 0.686$, 
$h=0.71$, $\sigma_8 = 0.84$.  Initial conditions 
are provided at $z=50$ by applying the \citet{klypin1997} transfer function 
and using the \citet{zeldovich1970} approximation. 
The simulation evolves 256$^3$ dark matter particles, 
with a mass resolution of $1.227 \times 10^{11} M_{\odot}$. 
Two comparison papers (\citet{heitmann05}, \citet{heitmann2008})
demonstrated good agreement on several derived quantities 
among ten different, widely used cosmology codes. 

\begin{figure}[t]
\begin{center}
    \includegraphics[width=140mm]{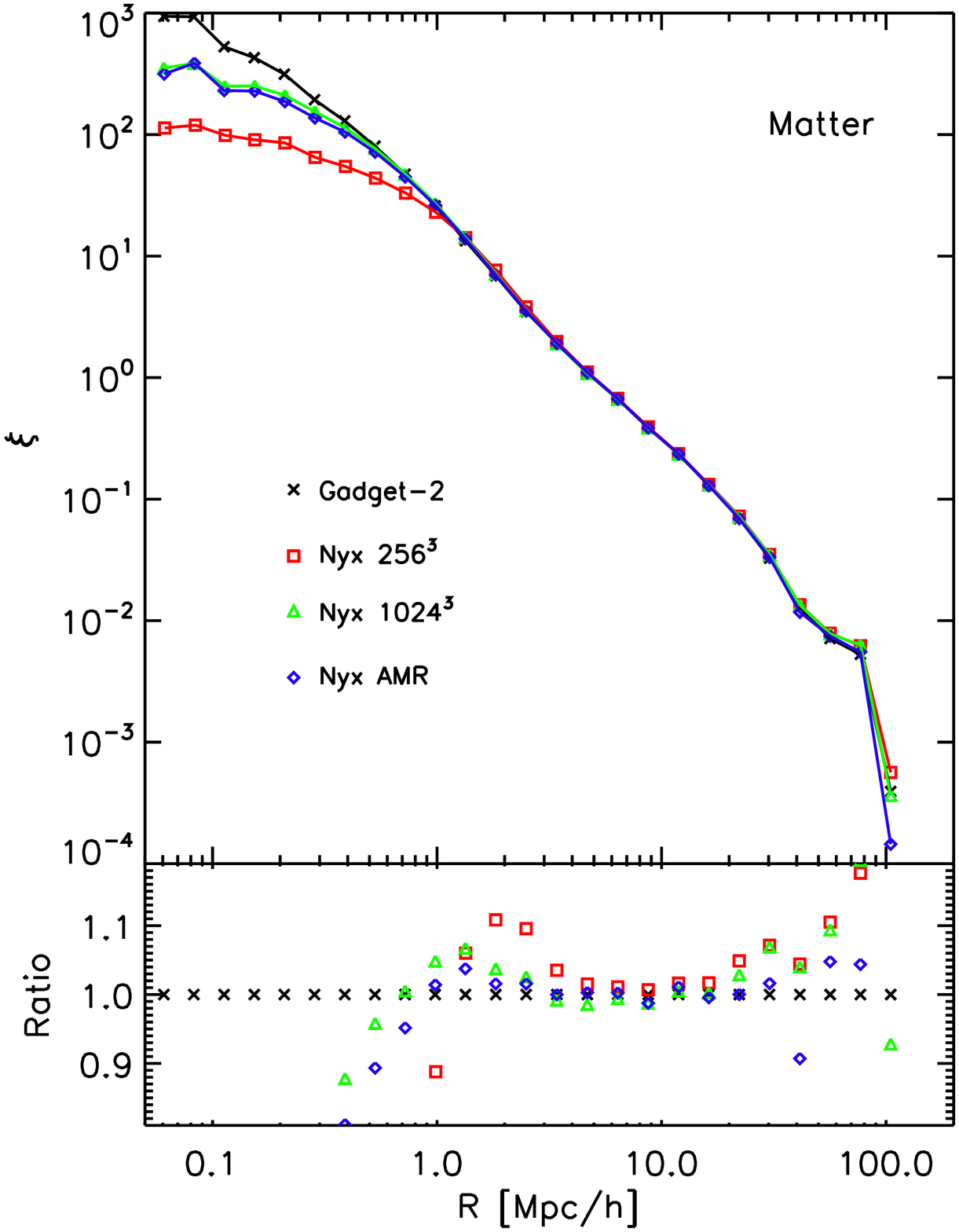} 
\end{center}
    \caption{Two-point correlation function of dark matter particles in 
     $256$ Mpc/h box (upper panel), and the ratio with respect to the Gadget-2 results (lower panel). 
     We show Nyx results from two uniform-grid simulations and one AMR simulation with a 
     256$^3$ base grid and two levels of refinement, each by a factor of two.} 
    \label{fig:dm_corr}
\end{figure}

Here we compare our results to the Gadget-2 simulations as presented in the Heitmann et al.~papers;
we refer interested readers to those papers to see a comparison of the Gadget-2 simulations with 
other codes.  We present results from three different Nyx simulations: uniform
grid simulations at 256$^3$ and 1024$^3$, and a simulation with a 256$^3$ base grid and two 
levels of refinement, each by a factor of two.
In Fig.~\ref{fig:dm_corr}, we show correlation functions (upper panel), and ratio to the Gadget-2 results 
(lower panel), to allow for a closer inspection. We see a strong match between the Gadget-2 and
Nyx results, and observe convergence of Nyx code towards the Gadget-2 results as the resolution
increases.  In particular, we note that the effective $1024^3$ results achieved with AMR very closely
match the results achieved with the uniform $1024^3$ grid.

\begin{figure}[t]
\begin{center}
    \includegraphics[width=140mm]{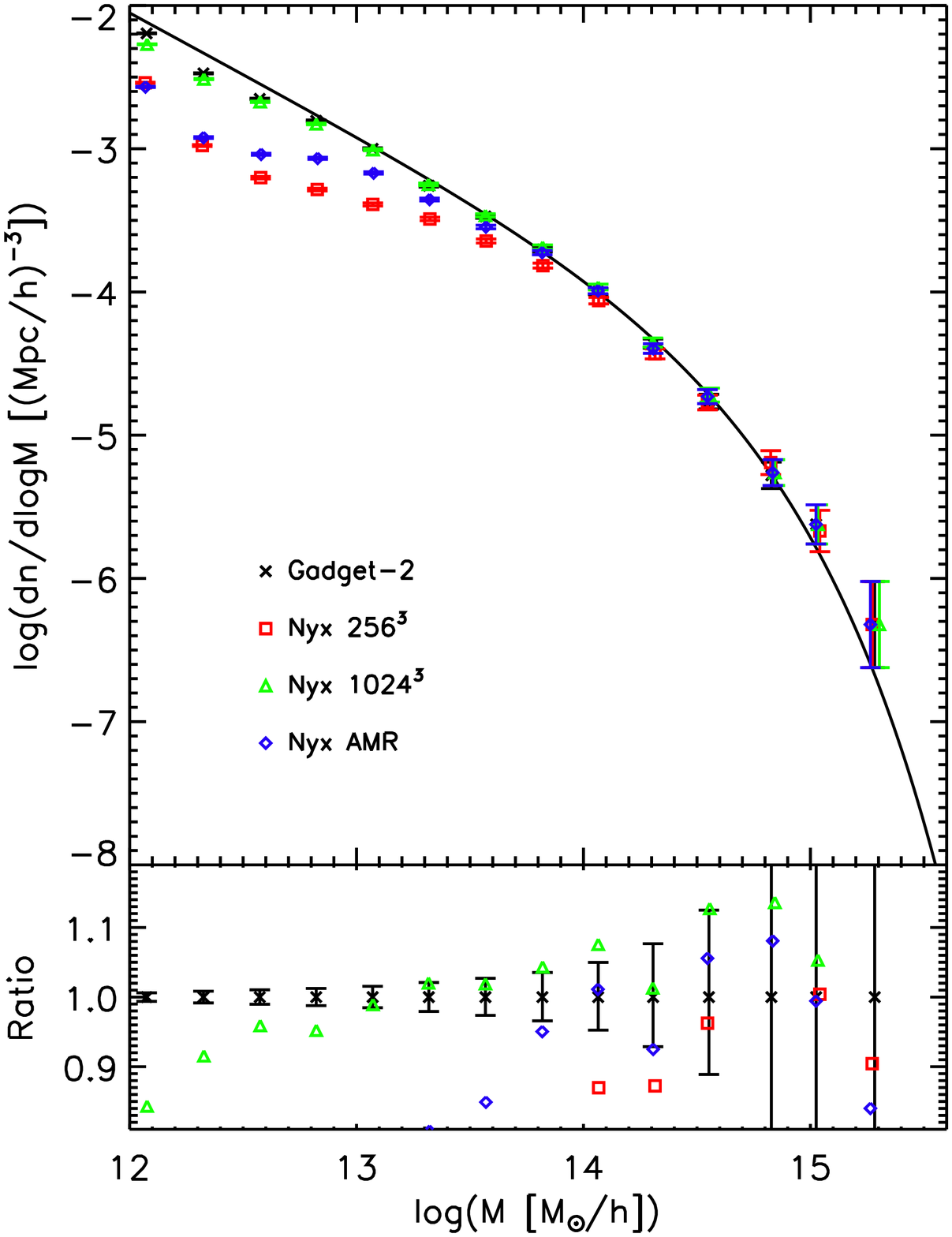} 
\end{center}
    \caption{Halo mass function also shows good convergence with the 
                    increasing force resolution towards Gadget-2 results, run at higher resolution. Solid line 
                    is \citep{sheth1999} fit, shown here purely to guide the eye; ratios are taken with respect to 
                    Gadget as in the other plots.} 
    \label{fig:dm_massf}
\end{figure}

Next, we examine the mass and spatial distribution of halos, two important statistical measures used 
in different ways throughout cosmology.
To generate halo catalogs we use the same halo finder for all runs, 
which finds friends-of-friends halos (FOF; \citet{davis1985}) using a linking length of $b=0.2$. 
To focus on the differences between codes we consider halos with as few as ten particles,
although in a real application we would consider only halos with at least hundreds of particles in 
order to avoid large inaccuracies due to the finite sampling of FOF halos. 
In Figure~\ref{fig:dm_massf}, we show results for the mass 
function, and we confirm good agreement between the high resolution Nyx run and Gadget-2. 
As shown in \citet{oshea2005}, \citet{heitmann05}, and \citet{lukic2007}, common refinement 
strategies suppress the halo mass function at the low-mass end. This is because small halos 
form very early and throughout the whole simulation domain; capturing them requires 
refinement so early and so wide, that block-based refinement (if not AMR in general) 
gives hardly any advantage over a fixed-grid simulation. Nyx AMR results shown here are 
fully consistent with the criteria given in \citet{lukic2007}, and are similar to other 
AMR codes \citep{heitmann2008}. 

\begin{figure*}[t]
\begin{center}$
\begin{array}{ccc}
    \includegraphics[width=66mm]{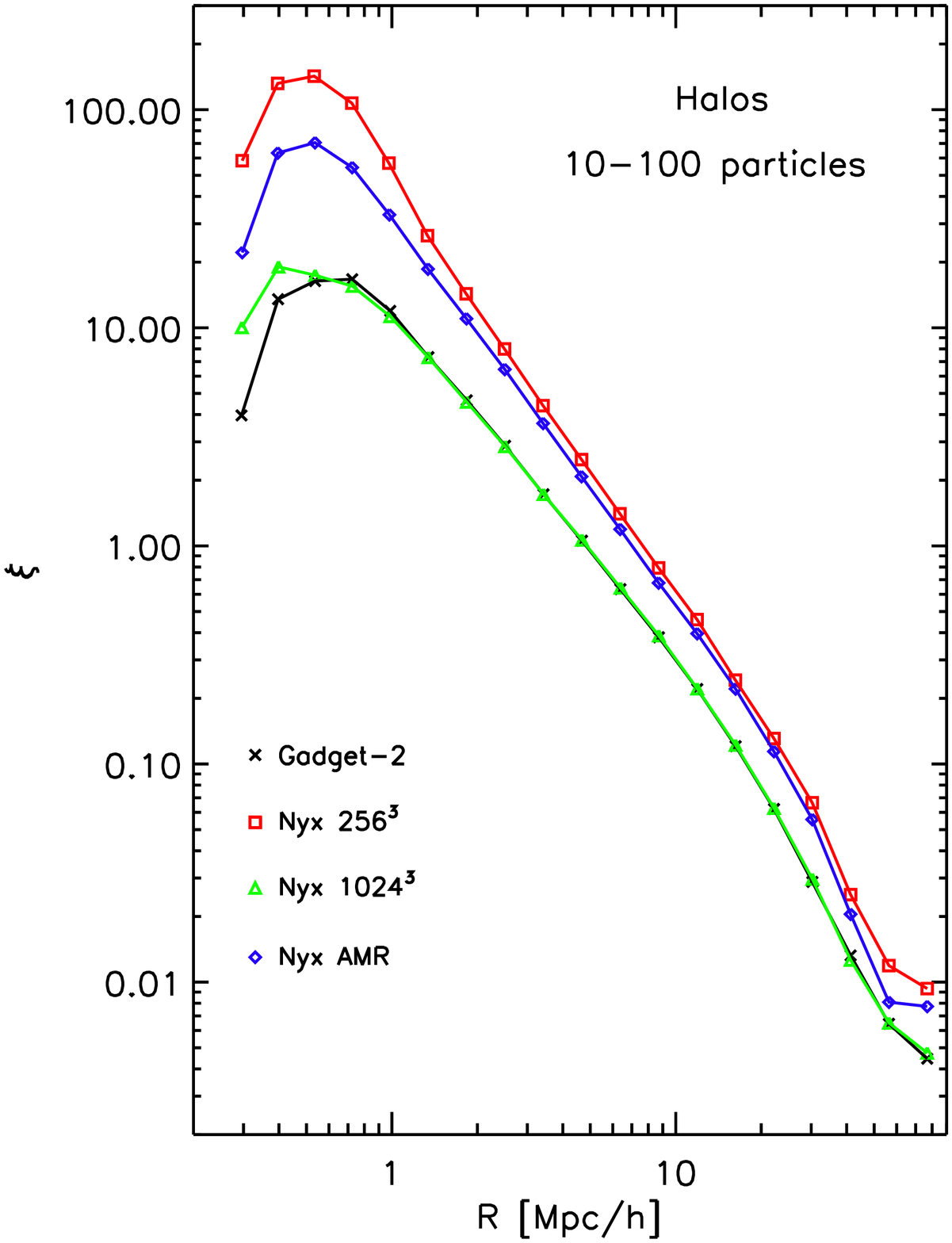} & 
    \hspace{-2.0cm} 
    \includegraphics[width=66mm]{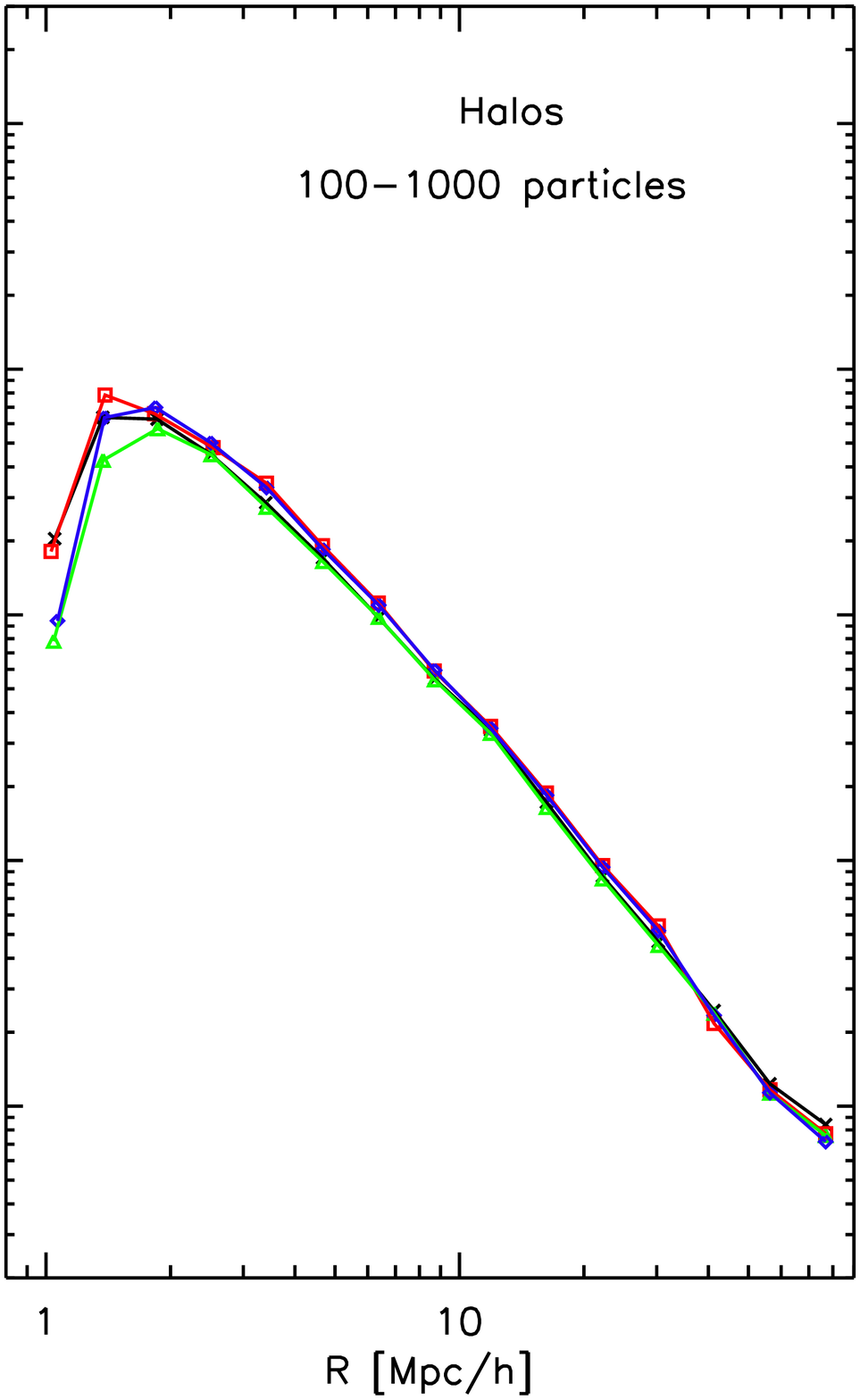} & 
    \hspace{-2.0cm} 
    \includegraphics[width=66mm]{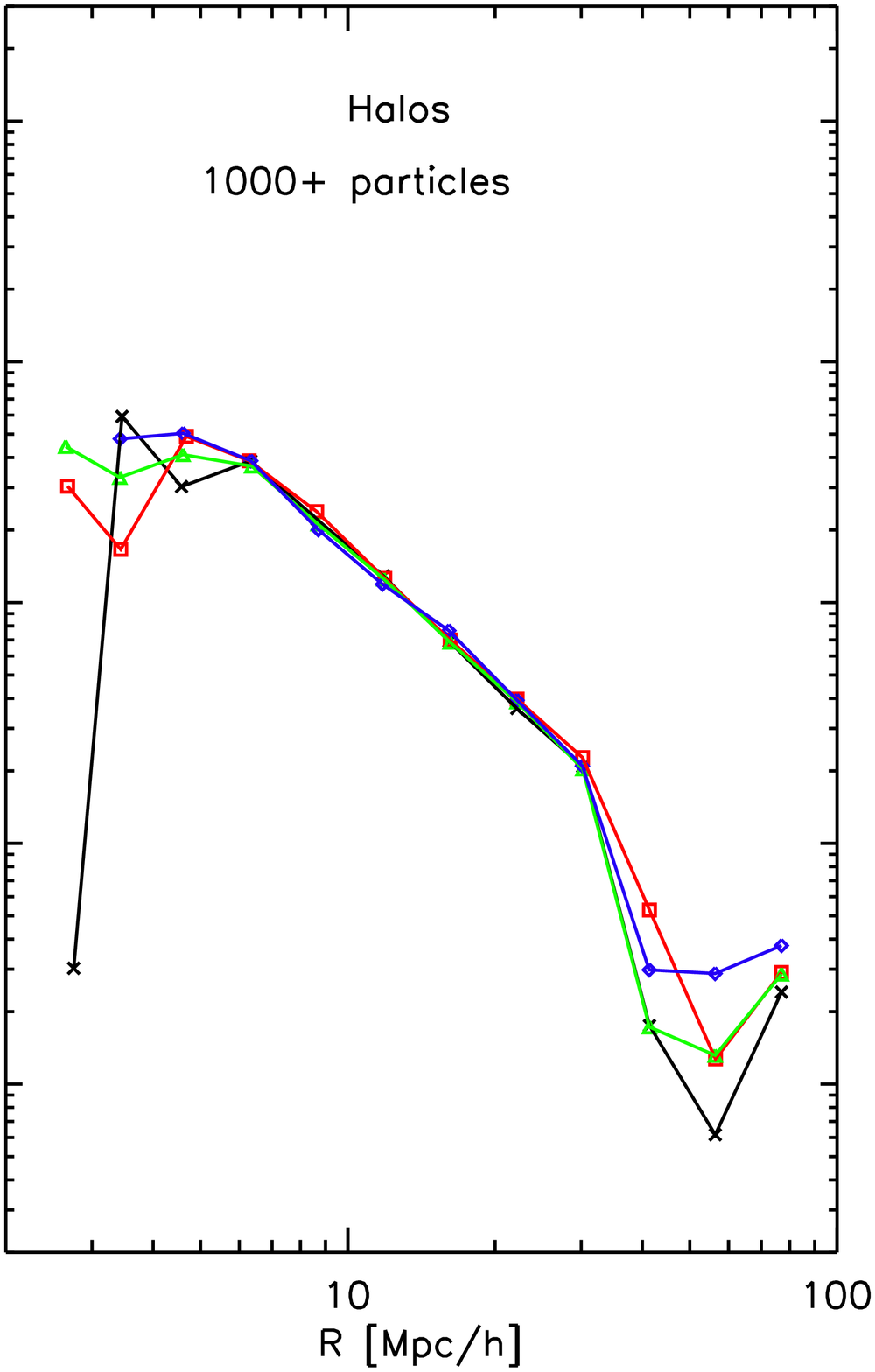}
\end{array}$
\end{center}
    \caption{Two-point correlation function of dark matter halos, separated in three mass groups. 
                    We see good agreement between the two codes, as well as rapid convergence even for halos sampled 
                    with a few tens of particles. We emphasize that 
                    halos in the range of 10--100 particles would not have been given serious consideration in a 
                    real application, but we show them here solely to examine differences between codes. 
                    Note that x-axis is different in all 3 panels. } 
    \label{fig:dm_corr_halos}
    \vspace{0.20 truecm}
\end{figure*}

\begin{figure*}[t]
\begin{center}$
\begin{array}{ccc}
    \includegraphics[width=66mm]{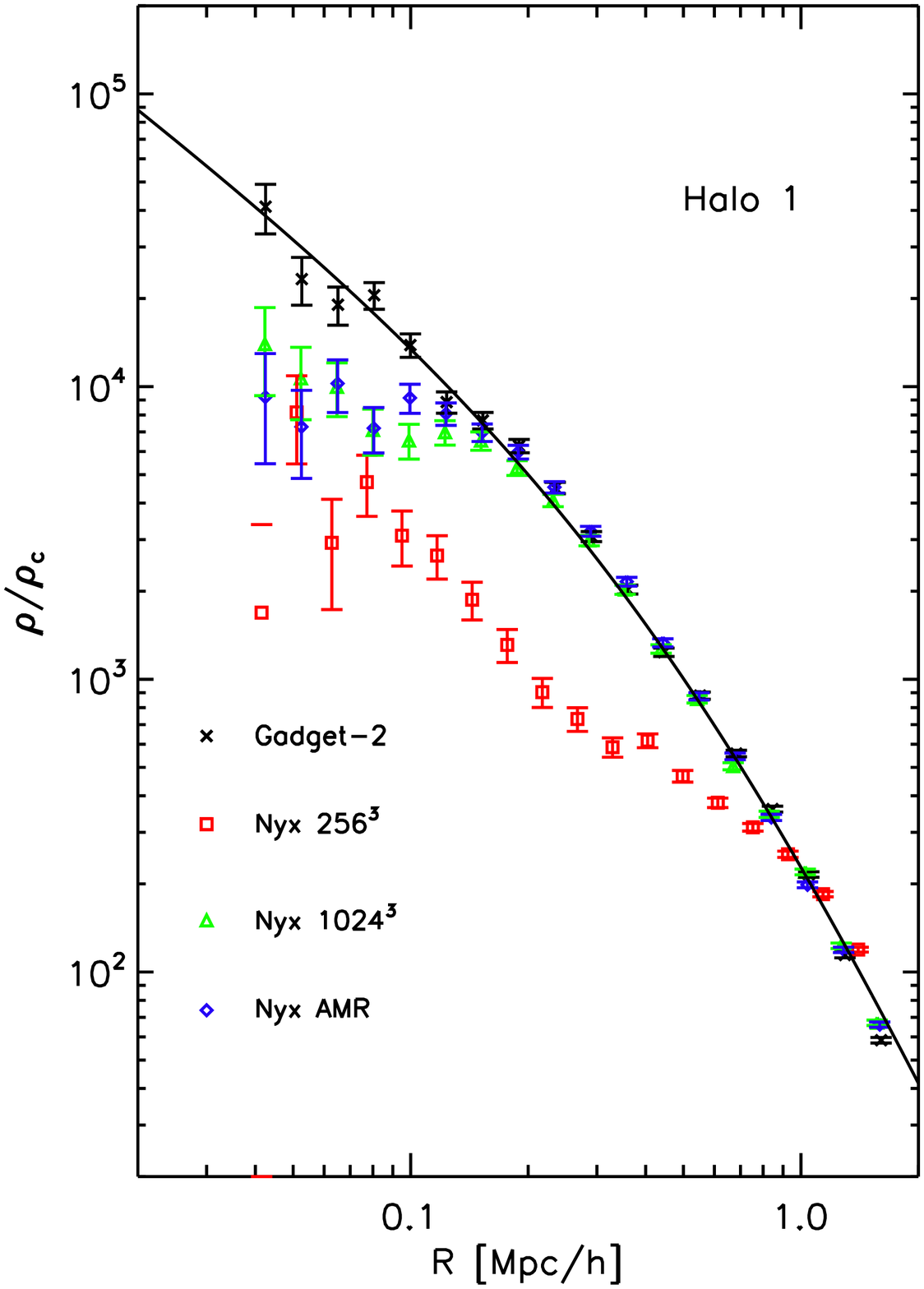} & 
    \hspace{-2.0cm} 
    \includegraphics[width=66mm]{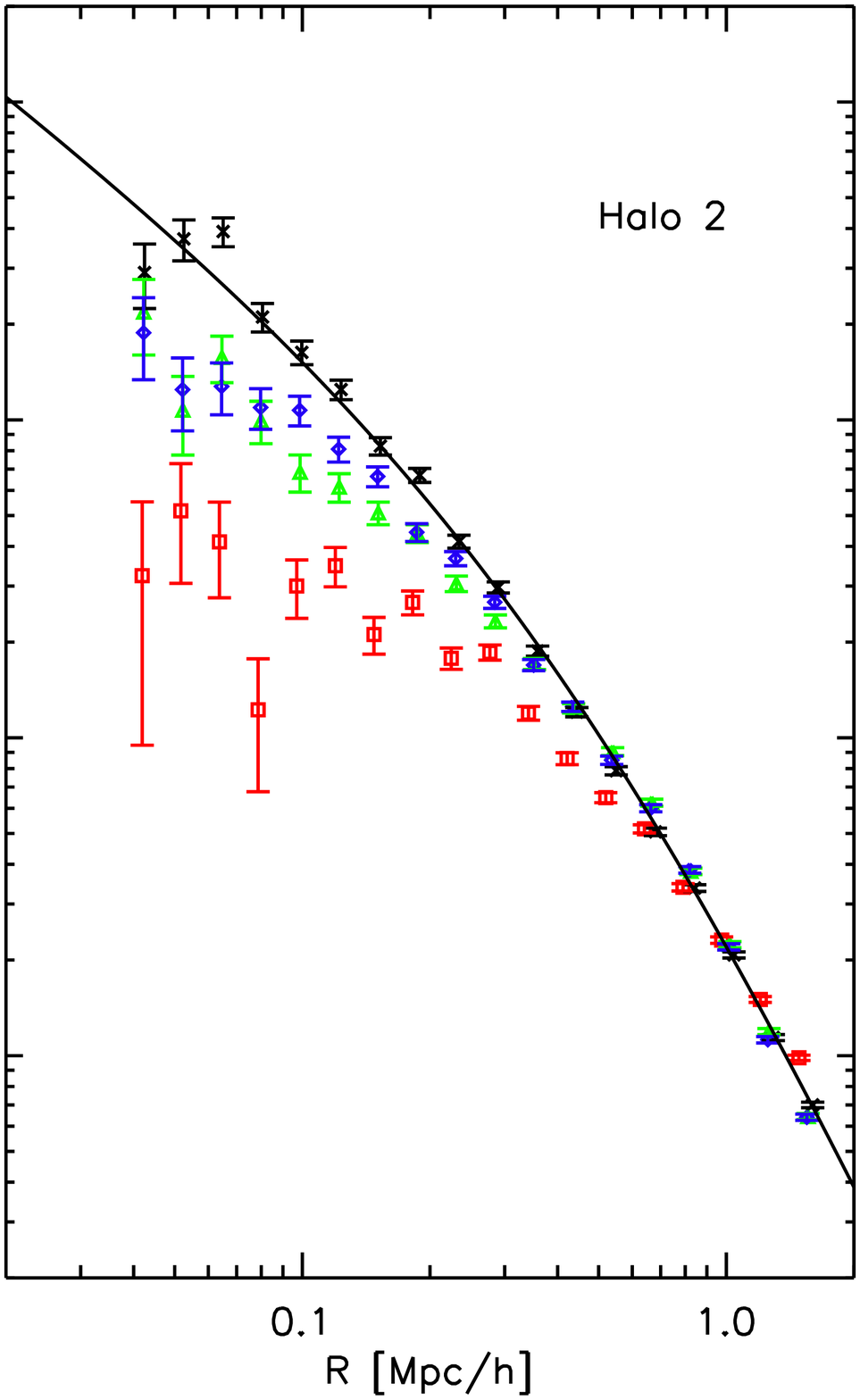} & 
    \hspace{-2.0cm} 
    \includegraphics[width=66mm]{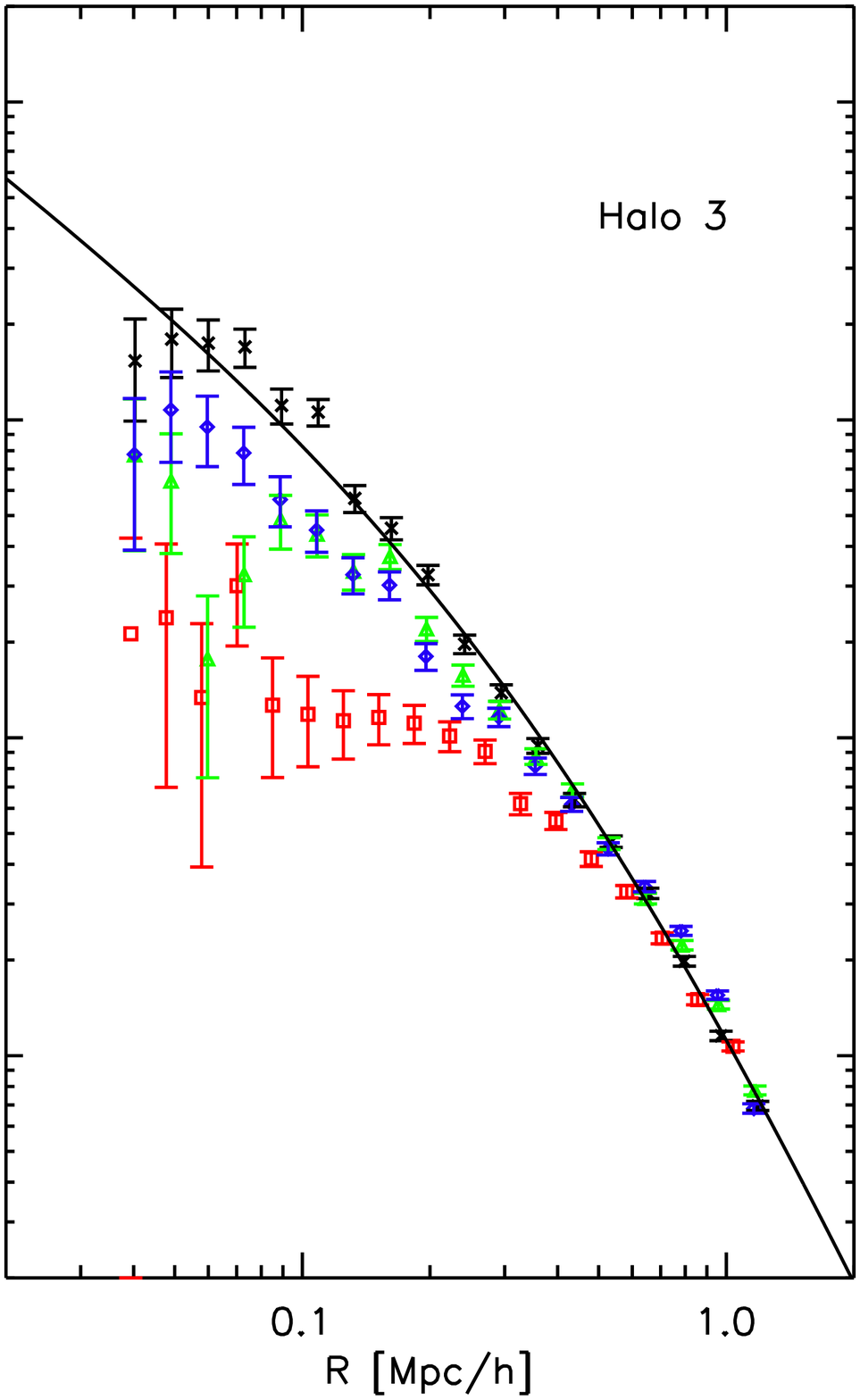}
\end{array}$
\end{center}
    \caption{Density profiles for the three most massive objects in the Gadget--2 run. Lines are the 
             best fit NFW profile to the Gadget--2 run. 
             We can see that the profiles from the AMR simulation closely match those of the uniform 1024$^3$ simulation.}
    \label{fig:halo_profiles}
    \vspace{0.20 truecm}
\end{figure*}

In Figure~\ref{fig:dm_corr_halos} we present the correlation function for halos. 
Rather than looking at all halos together, we separate them into three mass bins. 
For the smallest halos we see the offset in the low resolution run, but even that converges quickly, 
in spite of the fact that the 1024$^3$ run still has fewer small-mass halos than the Gadget run 
done with force resolution several times smaller (Fig.~\ref{fig:dm_massf}). 
Finally, in Figure~\ref{fig:halo_profiles} we examine the inner structure of halos.
We observe excellent agreement between the AMR run and the uniform $1024^3$ run, confirming the 
expectation that the structure of resolved halos in the AMR simulation match those in the fixed-grid runs. 
Overall, the agreement between the Nyx and Gadget-2 results is as expected, and is on a par 
with other AMR cosmology codes at this resolution as demonstrated in the Heitmann et al. papers.


\subsection{Santa Barbara Cluster Simulations}

\begin{figure}[t]
    \plotone{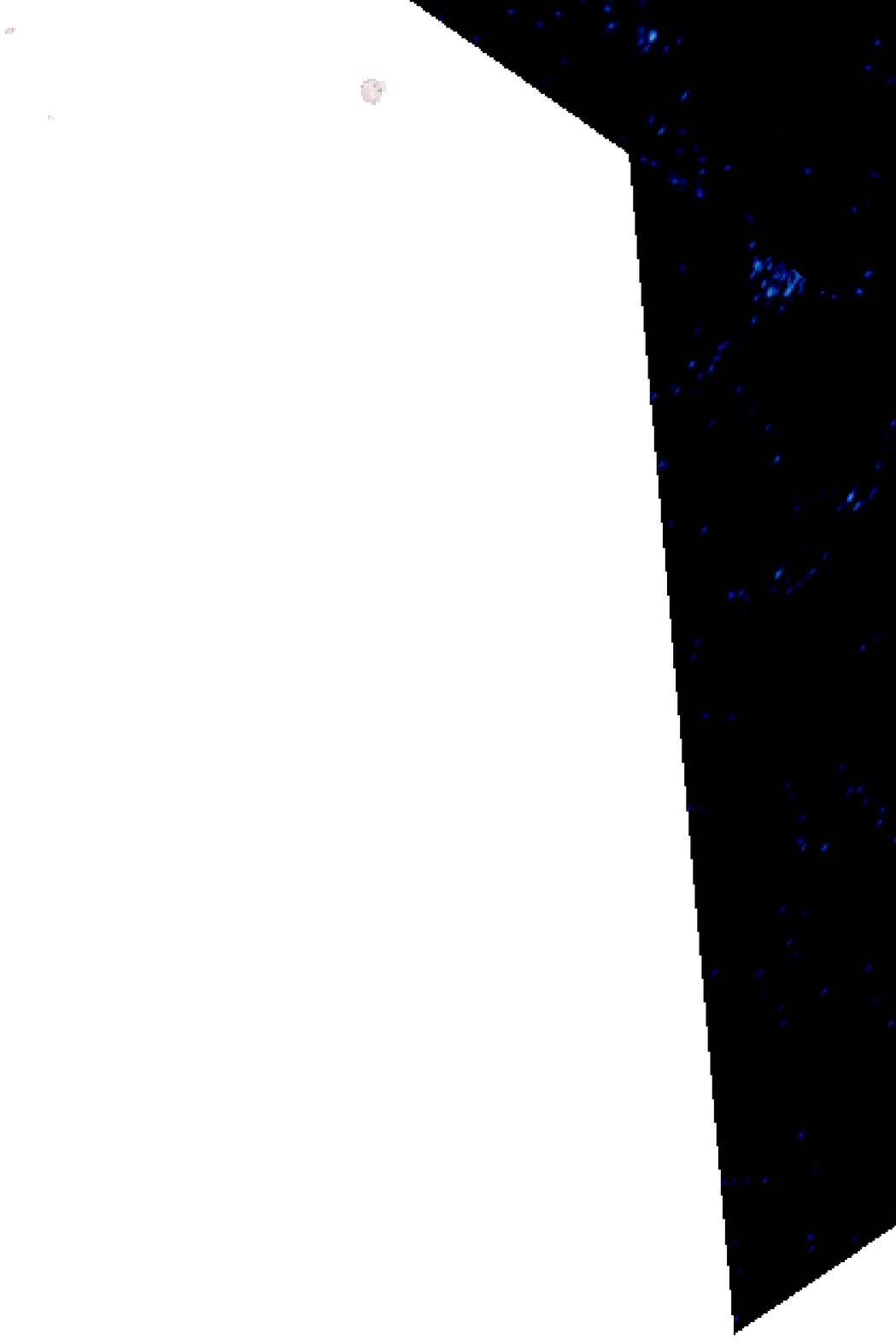}
    \caption{Santa Barbara cluster simulation. We show 3 orthogonal cuts through 
             the box center showing dark matter density in blue-white colors. 
             Superposed in red are iso-temperature surfaces of 10$^7$K gas showing rich 
             structure of intergalactic gas throughout the simulations domain.}
    \label{fig:SB_dm-T}
    \vspace{0.20 truecm}
\end{figure}

The Santa Barbara cluster has become a standard code comparison test problem for cluster formation 
in a realistic cosmological setting.  
To make meaningful comparisons, all codes start from the same set of initial conditions and evolve
both the gas and dark matter to redshift $z=0$ assuming an ideal gas equation of state 
with no heating or cooling (``adiabatic hydro'').
Initial conditions are constrained such that a cluster (3 $\sigma$ 
rare) forms in the central part of the box. 
The simulation domain is 64 Mpc on a side, with the SCDM cosmology: $\Omega = \Omega_m
= 1$, $\Omega_b = 0.1$, $h = 0.5$, $\sigma_8 = 0.9$, and $n_s = 1$. 
A 3D snapshot of the dark matter density and gas temperature at $z=0$ is shown in Figure~\ref{fig:SB_dm-T}.
For the Nyx simulation we use the exact initial conditions from \citet{heitmann05},  and begin
the simulation at $z = 63$ with $256^3$ particles. 
Detailed results from a code comparison study of the Santa Barbara test problem 
were published in \citet{frenk1999}; we present here not all, but the
most significant diagnostics from that paper for comparison of Nyx with other available codes.
We refer the interested reader to that paper for full details of the simulation results for
the different codes.

We first examine the global cluster properties, calculated inside $r_{200}$ with respect to the 
critical density at redshift $z=0$. We find the total mass of the cluster to be 
$1.15 \times 10^{15} M_{\odot}$; \citet{frenk1999} report 
the mean and 1 $\sigma$ deviation as $(1.1 \pm 0.05) \times 10^{15} M_{\odot}$ for all codes. 
The radius of the cluster from the Nyx run is 2.7 Mpc; \citet{frenk1999}
report $2.7 \pm 0.04$. The NFW concentration is 7.1; the original paper gives 7.5 as a rough guideline to 
what the concentration should be.

\begin{figure*}[t]
\begin{center}$
\begin{array}{cc}
    \hspace{-1.5 cm}
    \includegraphics[width=76mm]{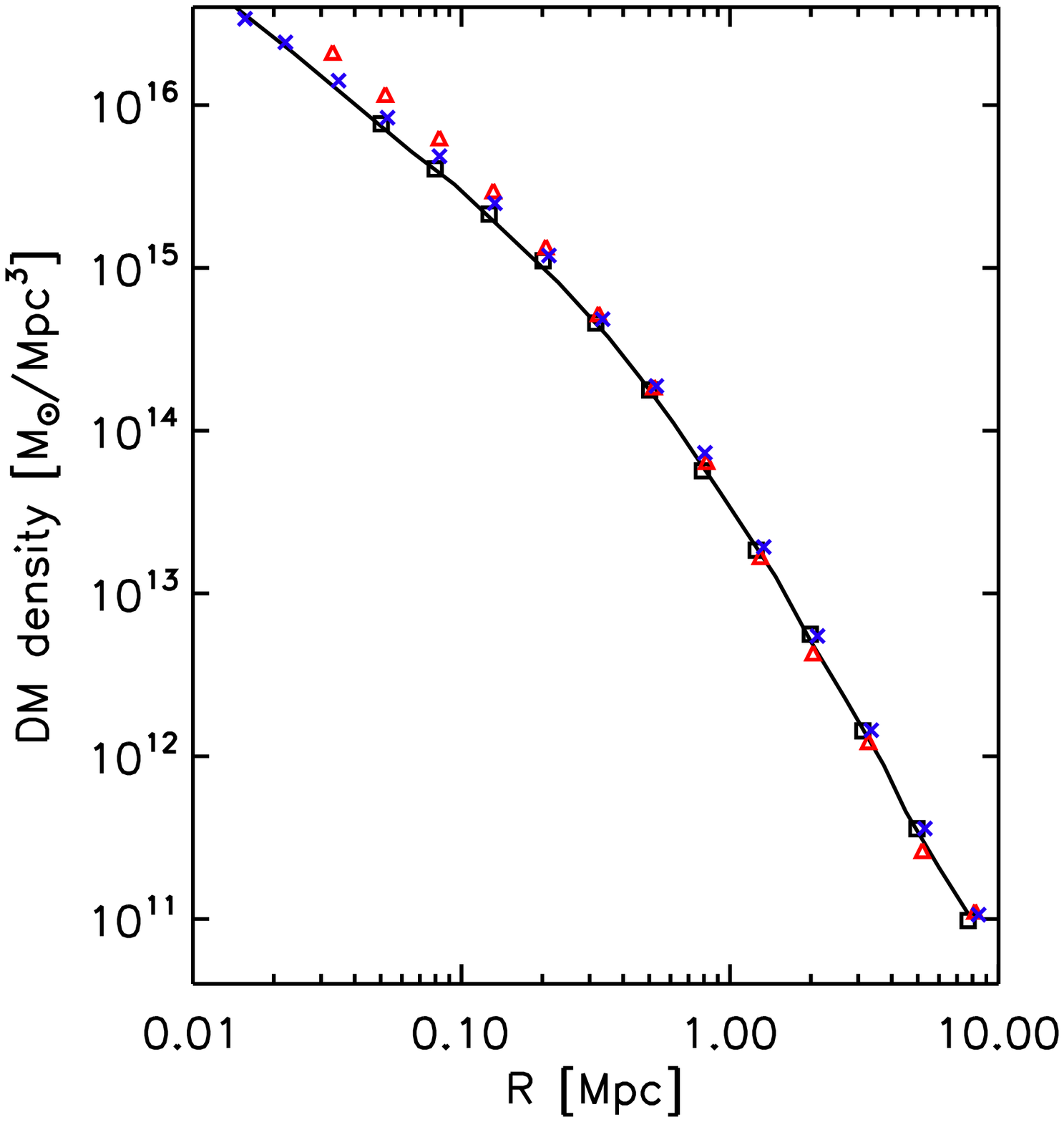} & 
    \includegraphics[width=76mm]{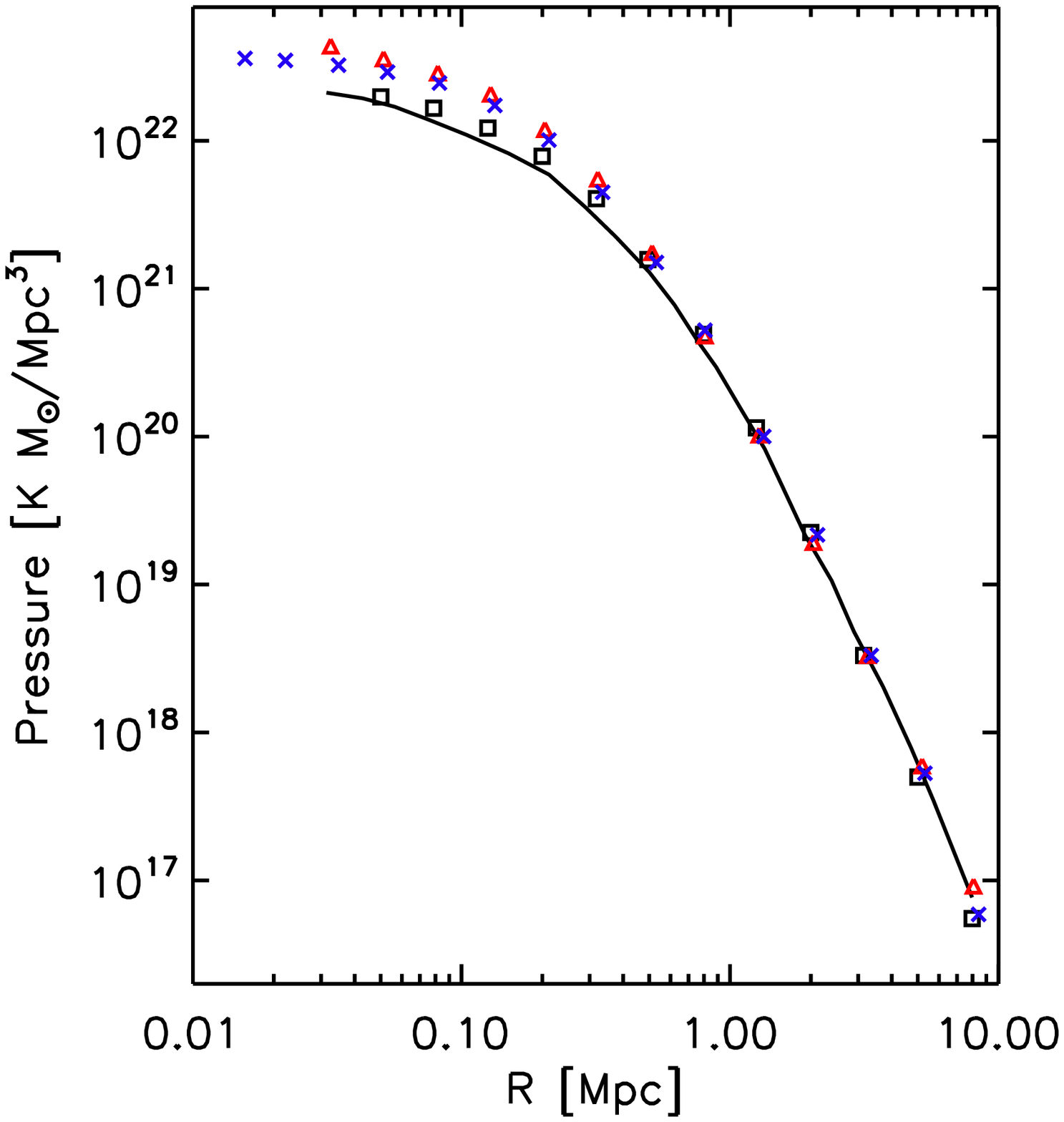} \\ 
    \hspace{-1.5 cm}
    \includegraphics[width=76mm]{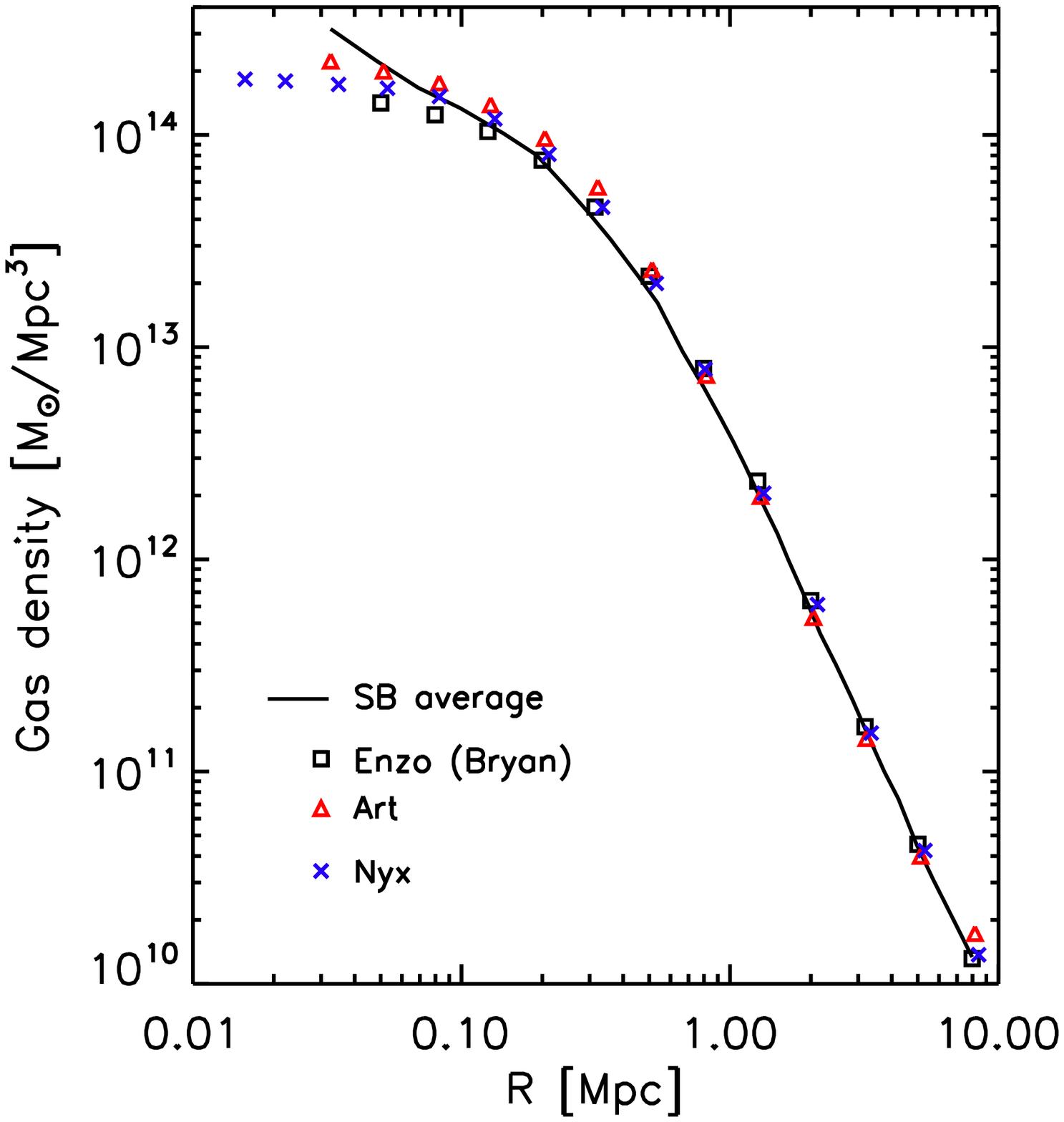} & 
    \includegraphics[width=76mm]{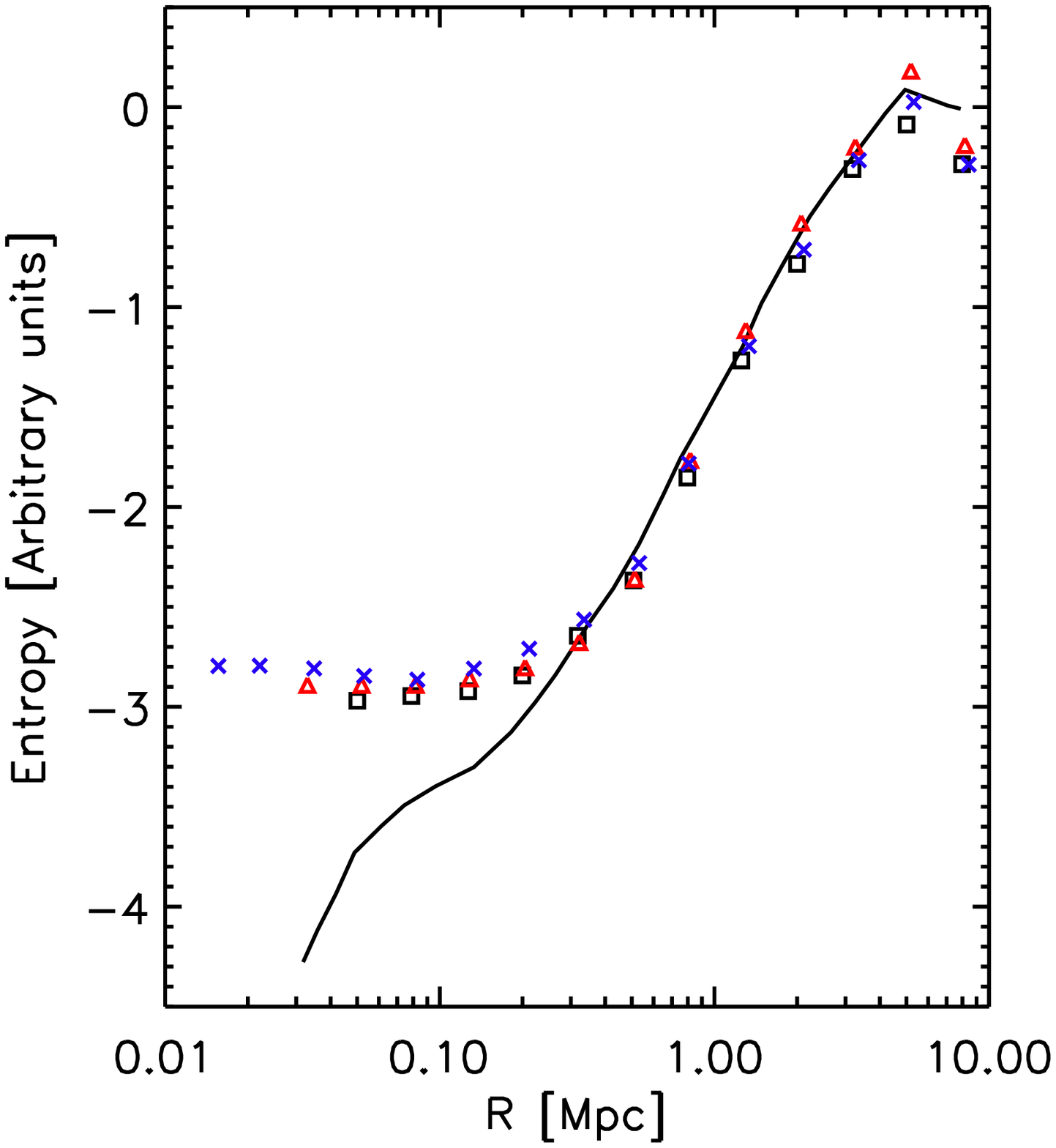}  
\end{array}$
\end{center}
    \caption{Left: density of dark matter (upper panel) and gas (lower panel); 
             right: pressure (upper) and entropy (lower) profiles for the
             Santa Barbara cluster. Besides Nyx, we show the average from 
             the original Frenk et al. 1999 paper, and results from 
             two AMR codes, Enzo and ART.}
    \label{fig:SB_profiles}
    \vspace{0.20 truecm}
\end{figure*}

In Figure~\ref{fig:SB_profiles} we show comparisons of the radial profiles of several quantities:  
dark matter density, gas density, pressure, and entropy.  Since the original data from the comparison 
project is not publicly available,  we compare only with the average values, 
as well as the Enzo and ART data as estimated from the figures in \citet{kravtsov2002}.
We also note that we did not have exactly the same initial conditions as in 
\citet{frenk1999} or \citet{kravtsov2002} and our starting redshift 
differs from that of Enzo ($z=30$), so we do not expect exact agreement.

In the original code comparison, the 
definition of the cluster center was left to the discretion of each simulator, and here we adopt 
the gravitational potential minimum to define the cluster center. As in 
\citet{frenk1999}, we radially bin 
quantities in 15 spherical shells, evenly spaced in log radius, and covering 3 orders of magnitude -- 
from 10 kpc to 10 Mpc. The pressure reported is $p=\rho_{gas}T$, and entropy is 
defined as $s = \ln(T/\rho_{gas}^{2/3})$. We find good agreement between Nyx and the other AMR codes, 
including the well-known entropy flattening in the central part of the cluster. 
In sharp contrast to grid methods, SPH codes find central entropy to continue rising towards smaller radii. 
\cite{Mitchell2009} made a detailed investigation of this issue, and have found that the difference 
between the two is independent of mesh sizes in grid codes, or particle number in SPH. Instead, the 
difference is fundamental in nature, and is shown to originate as a consequence of the suppression of eddies 
and fluid instabilities in SPH. 
(See also \citet{abel2011} for an SPH formulation that improves mixing at a cost 
of violating exact momentum conservation.)   Overall, we find excellent agreement
of Nyx with existing AMR codes.

\section{Conclusions and Future Work}\label{sec:con}

We have presented a new N-body and gas dynamics code, Nyx, 
for large-scale cosmological simulations.   Nyx is designed to 
efficiently utilize tens of thousands of processors; timings of the code 
up to almost 50,000 processors show excellent weak scaling behavior.   
Validation of Nyx in pure dark matter runs and dark matter with adiabatic
hydrodynamics has been presented.  Future papers will give greater detail
on the implementation in Nyx of source terms, and will present results 
from simulations incorporating different heating and cooling mechanisms of the gas, 
as needed for increased fidelity in different applications.  Scientific studies already 
underway with Nyx include studies of the Lyman-$\alpha$ forest and galaxy clusters. 
In addition, we plan to extend Nyx to allow for simulation of alternative cosmological 
models to $\Lambda$CDM, most interestingly dynamical dark energy and 
modifications of Einstein's gravity. The existing grid structure and multigrid Poisson solver 
should make it straightforward to extend the current capability to 
iterative methods for solving the non-linear elliptic equations arising in those models. 


\acknowledgements

This work was supported by Laboratory Directed Research and
Development funding (PI: Peter Nugent) from Berkeley Lab, 
provided by the Director, Office of Science, of the 
U.S. Department of Energy under Contract No. DE-AC02-05CH11231.  
Additional support for improvements to BoxLib to support the Nyx code and others 
was provided through the SciDAC FASTMath Institute, funded by the 
Scientific Discovery through Advanced Computing (SciDAC) program funded 
by U.S. Department of Energy Office of Advanced Scientific Computing Research 
(and Office of Basic Energy Sciences/Biological and 
Environmental Research/High Energy Physics/Fusion Energy Sciences/Nuclear Physics).
Calculations presented in this paper used resources of the National Energy Research Scientific
Computing Center (NERSC), which is supported by the Office of Science of the
U.S. Department of Energy under Contract No. DE-AC02-05CH11231.

We thank Peter Nugent and Martin White for many useful discussions.

\pagebreak
\bibliographystyle{apj}
\bibliography{refs}


\end{document}